\documentclass[a4paper,fleqn,usenatbib]{mnras}

\usepackage{newtxtext,newtxmath}

\usepackage[T1]{fontenc}
\usepackage{ae,aecompl}

\usepackage{graphicx}	
\usepackage{amsmath}	
\usepackage{amssymb}	
\usepackage{mathtools,bm}

\usepackage{hyperref}
\hypersetup{
  colorlinks=true,        
  linkcolor=blue,         
  citecolor=cyan,         
  filecolor=magenta,      
  urlcolor=magenta            
 }

\usepackage{graphicx}
\usepackage{color}
\usepackage{soul}
\definecolor{orange}{rgb}{1,0.5,0}

\newcommand{\cf}{cf.,~}
\newcommand{\ie}{i.e.,~}
\newcommand{\eg}{e.g.,~}

\title[Two-moment scheme for general-relativistic radiation
  hydrodynamics: a systematic description]
  {Two-moment scheme for general-relativistic radiation
  hydrodynamics: a systematic description and new applications}

\author[L.R. Weih et al.]{Lukas R. Weih$^{1}$\thanks{weih@itp.uni-frankfurt.de}, 
Hector Olivares$^{2,1}$, Luciano Rezzolla$^{1,3,4}$
\\
$^{1}$ Institut f\"ur Theoretische Physik, Goethe Universit\"at Frankfurt am Main, Germany\\
$^{2}$ Department of Astrophysics/IMAPP, Radboud University Nijmegen, The Netherlands\\
$^{3}$ School of Mathematics, Trinity College, Dublin 2, Ireland\\
$^{4}$ Helmholtz Research Academy Hesse for FAIR, Max-von-Laue-Str. 12, 60438 Frankfurt, Germany}

\date{Accepted XXX. Received YYY; in original form ZZZ}

\pubyear{2020}

\begin{document}
\label{firstpage}
\maketitle

\begin{abstract}
  We provide a systematic description of the steps necessary -- and of
  the potential pitfalls to be encountered -- when implementing a
  two-moment scheme within an Implicit-Explicit (IMEX) scheme to
  include radiative-transfer contributions in numerical simulations of
  general-relativistic (magneto-)hydrodynamics. We make use of the M1
  closure, which provides an exact solution for the optically thin and
  thick limit, and an interpolation between these limits. Special
  attention is paid to the efficient solution of the emerging set of
  implicit conservation equations. In particular, we present an
  efficient method for solving these equations via the inversion of a
  $4\times 4$-matrix within an IMEX scheme. While this method relies on
  a few approximations, it offers a very good compromise between
  accuracy and computational efficiency. After a large number of tests
  in special relativity, we couple our new radiation code,
  \texttt{FRAC}, with the general-relativistic magnetohydrodynamics
  code \texttt{BHAC} to investigate the radiative Michel solution,
  namely, the problem of spherical accretion onto a black hole in the
  presence of a radiative field. By performing the most extensive
  exploration of the parameter space for this problem, we find that the
  accretion's efficiency can be expressed in terms of physical
  quantities such as temperature, $T$, luminosity, $L$, and black-hole
  mass, $M$, via the expression $\varepsilon=(L/L_{\rm
  Edd})/(\dot{M}/\dot{M}_{\rm Edd})= 7.41\times
  10^{-7}\left(T/10^6\,\mathrm{K}\right)^{0.22}
  \left(L/L_\odot\right)^{0.48} \left(M/M_\odot\right)^{0.48}$, where
  $L_{\mathrm{Edd}}$ and $\dot{M}_{\mathrm{Edd}}$ are the Eddington
  luminosity and accretion rate, respectively. Finally, we also
  consider the accretion problem away from spherical symmetry, finding
  that the solution is stable under perturbations in the radiation
  field.
\end{abstract}


\section{Introduction}
\label{sec:intro}

The study of high-energy astrophysical phenomena plays an increasingly
important role in understanding the fundamental laws of the universe.
One reason for this was the beginning of the multimessenger era, which
was ushered in by the first detection of gravitational waves of a binary
neutron star system by the LIGO/VIRGO collaboration \citep{Abbott2017}
and its electromagnetic counterpart
\citep{Abbott2017b,Abbott2017d_etal}. This event, GW170817, provided a
wealth of information not just on the nature of gravity, but also on the
properties of matter under extreme conditions \citep[see][for an
  incomplete
  list]{Margalit2017,Bauswein2017b,Rezzolla2017,Ruiz2017,Annala2017,Radice2017b,Most2018,Coughlin2018a,Burgio2018,Tews2018,Shibata2019,Koeppel2019}. Another
milestone for high-energy astrophysics were the millimetre-wavelength
observations by the Event Horizon Telescope (EHT) collaboration, which
delivered the first spatially resolved image of a black-hole shadow in
the center of the galaxy M87
\citep{EHT_M87_PaperI,EHT_M87_PaperV,EHT_M87_PaperVI}. This image shows
an asymmetric emission ring around the central black hole, which can be
explained by the model of a Kerr black hole within the context of general
relativity.

Both of these recent milestones have in common that the understanding of
these observations was aided by numerical simulations, either of binary
neutron stars \citep[see, \eg][for reviews]{Baiotti2016, Paschalidis2016}
or the accretion onto black holes \citep[see, \eg][for a
  review]{Abramowicz2013}. These systems are nowadays simulated by
solving numerically the equations of general-relativistic
magnetohydrodynamics (GRMHD). However, as the realism of these
simulations increases, it becomes necessary to include the coupling
between the fluid describing the neutron star or the accretion disk with
radiation in the form of photons or neutrinos, resulting in what is
called general-relativistic radiative-transfer MHD (GRRTMHD). In standard
simulations of geometrically thick, optically thin accretion disks, this
coupling is less important and it is therefore adequate to neglect the
backreaction onto the fluid of the radiation, which can instead be
handled independently and in a post-processing stage
\citep{Mizuno2018,EHT_M87_PaperV,Davelaar2019}. Although neglecting the
interaction between fluid and the radiation field is a good approximation
for a low-luminosity active galactic nuclei (LLAGNs) such as Sgr~A*, the
inclusion of radiative cooling has been recently considered to produce
self-consistent models of another LLAGN such as M87
\citep{Moscibrodzka2011,Dibiet12}. Detailed simulations of this source
including radiation interaction were carried out recently by
\citet{Chael2019}. On the other hand, in systems of compact objects with
high accretion rates, \ie close to or above their Eddington limit, the
disk cools efficiently via the production of photons that are then
radiated to infinity. In such radiation-dominated accretion flows, the
dynamical interaction between radiation and fluid becomes non-negligible
\citep[see][and reference therein]{McKinney2014}. Also for binary
neutron-star simulations, and especially during the post-merger phase,
the dynamical evolution of radiation in the form of neutrinos and the
full coupling to the fluid is necessary. Indeed, after the merger of two
neutron stars, the composition and amount of the ejected material can be
significantly altered due to interactions of the fluid with neutrinos
that are produced within the hot merger remnant. The properties of this
material are directly connected to the resulting kilonova, which results
from the radioactive decay of elements produced via r-process in the
ejected material \citep{Rosswog2014a,Dietrich2016, Siegel2016a,
  Bovard2017, Perego2017, Fujibayashi2017b, Siegel2017, Fernandez2018,
  Most2019b}. It is thus necessary to solve the equations of radiative
transport in conjunction with those describing the dynamics of the fluid.

The equation describing the evolution of the radiation field is given by
the Boltzmann equation \citep[see, \eg][]{Rezzolla_book:2013}, which is
seven-dimensional (7D), since it has to evolve in time (one dimension)
variables defined both in the spatial space (three dimensions) and in the
momentum space (three dimensions). In contrast to the four-dimensional
(4D) equations of GRMHD, the numerical cost for solving the full
Boltzmann equation is prohibitive. Therefore, many approximate schemes
have been developed over the years. The most basic scheme is the
so-called ``leakage-scheme'', which only considers cooling of the fluid
via the emission of the radiation
\citep{Ruffert1996c,Rosswog:2003b,Galeazzi2013,Perego2014}, while heating
through absorption is neglected. A more accurate, yet still approximate
and feasible, approach is provided by the moment scheme. Within this
scheme, which is based on Thorne's moment formalism \citep{Thorne1981} --
and first implemented within general relativity by
\citet{Rezzolla1994,Shibata2011,Cardall2013b} -- only the first few
moments of the radiation distribution function are evolved. Within this
formalism, the lowest-order approximation is then represented by the
evolution of only the zeroth moment, and is often referred to as the
\textit{flux-limited diffusion} limit [\citet{Pomraning1981,
    Levermore1981}; see also \citet{Rahman2019} for an implementation of
  this scheme]. This approximation is particularly suited for spherically
symmetric problems, because it does not provide any information about the
direction of the radiation fluxes. This can be achieved, however, when
evolving also the first moment (momentum density) of the distribution
function in what is also called the ``M1 scheme''. While this choice
introduces three more variables to be evolved in 3D (the energy flux is a
three-vector), the M1 scheme offers the best compromise between accuracy
and feasibility and has widely been used in the context of black-hole
accretion
\citep{Zanotti2011,Fragile2012,Roedig2012,Sadowski2013,Fragile2014,McKinney2014},
core-collapse supernovae \citep{Oconnor2015,Just2015b,Kuroda2016},
black-hole--neutron-star mergers \citep{Foucart2015a,Foucart2016a} and
binary neutron stars \citep{Foucart2015,Sekiguchi2016}. Because of the
related very high computational costs, more accurate methods -- such as
the Monte-Carlo scheme \citep{Foucart2017,Miller2019a} -- have so far
been considered only in the post-processing of a binary neutron-star
simulation \citep{Foucart2018} or during the post-merger phase with a
fixed spacetime \citep{Miller2019b}.

We here present a detailed description of our implementation of the M1
scheme within the stand-alone Frankfurt Radiation Code, \texttt{FRAC},
that can easily be coupled to already existing GRMHD codes, either in
fixed spacetimes or in arbitrary and dynamically evolving
spacetimes. Several different implementations of the M1 scheme can be
found in the literature and to guide the reader in this rather ample
literature we note that the biggest differences among the various codes
can be restricted to three main aspects, which will be discussed in
detail throughout this work:
\begin{itemize}
\item the type of closure that determines which limit (optically thin
  and/or thick) can be treated (see Sec. \ref{sec:closure} for details).
\item the treatment of the radiative-transfer equations in the stiff
  limit. Here, we make use of an IMEX scheme \citep{Pareschi2010} in
  order to assure numerical stability also in the optically thick regime
  (see Sec. \ref{sec:implicit} for details). Keeping in mind the high
  computational cost for binary neutron-star simulations, we present an
  efficient way to solve the implicit equations of the IMEX scheme, which
  represents a good compromise between accuracy and computational cost.
\item the inclusion of a dependence on the frequency of the
  radiation. The evolved moments, in fact, depend not just on space and
  time, but also on the frequency of the photon/neutrino (see
  Sec. \ref{sec:evolution}). The inclusion of this additional dependency
  drastically increases the computational cost and has been so far
  considered only in few cases, as, \eg in the one-dimensional code of
  \citet{Oconnor2015}. For simplicity, no frequency dependence is
  considered here.
\end{itemize}

Besides the obvious presentation of a large set of tests in special
relativity, there are two important aspects in which our work here
differs from those presented so far in the literature. First, we provide
a rather detailed description of the numerical issues and problems that
had to be faced and solved when implementing the M1 scheme in a generic
general-relativistic MHD context, both in stationary and analytic
spacetime, but also within codes employed to simulate binary neutron-star
mergers. We hope that, in this way, many of the unexpected pitfalls we
have encountered and that were not documented before, can be easily
avoided by those wanting to replicate our results. Second, we consider as
a rather stringent test of our approach in a curved spacetime a problem
that actually has an astrophysical application, deriving an expression
that could be of interest in astronomy.

More specifically, we consider the problem of a spherically symmetric
accretion flow onto a black hole. While closed-form solutions are present
in the absence of radiation \citep{Bondi52, Michel72}, this scenario can
only be solved within a general-relativistic radiative-transfer (GRRT)
context when the ordinary fluid is coupled non-trivially with a radiation
fluid\footnote{Magnetic fields could in principle be introduced, but
  would imply a rather artificial scenario involving a monopolar magnetic
  field and an arbitrary strength}. Indeed, this is a classical GRRT
problem, which has been studied in the past
\citep{Vitello1978,Nobili1991} and more recently
\citep{Fragile2012,Roedig2012,Sadowski2013,Fragile2014,McKinney2014}. Here,
we explore the largest space of parameters characterising this problem
and obtain in this way a simple and useful relation between the accretion
efficiency and the black-hole's bolometric luminosity and mass. Such an
expression allows one, therefore, to simply relate observable quantities,
such as the luminosity and the temperature of the infalling fluid, to the
mass of the black hole. Finally, we also simulate this problem away from
spherical symmetry via introducing perturbations in the in the accreting
flow and hence in the radiation field. Our simulations show that the
accretion flow is stable and returns to its equilibrium after radiating
to infinity the excess energy introduced by the perturbation. This result
complements the interesting investigation of the Michel solution recently
performed by \citet{Tejeda2020,Waters2020} and which indicates that the
dynamics of this scenario is richer than what expected so far.

The paper is organised as follows: in Sec. \ref{sec:moment-scheme} we
list the equations of the truncated moment formalism and discuss the
details of our implementation, including the closure
(Sec. \ref{sec:closure}), the computation of the fluxes
(Sec. \ref{sec:fluxes}) and the IMEX scheme (Sec.
\ref{sec:implicit}). We then show the validity of our implementation with
a number of standard-tests in Sec. \ref{sec:tests}. After verifying the
correct coupling of \texttt{FRAC} with the ``Black Hole Accretion Code''
(\texttt{BHAC}) \citep{Porth2017,Olivares2019}
in Sec. \ref{sec:fluid-coupling}, we
finally apply the coupled code to the problem of spherically symmetric
accretion onto a black hole in Sec. \ref{sec:bondi}. In
Sec. \ref{sec:perturbation} we present the solution of this problem when
deviating from spherical symmetry and finally conclude and summarizes in
Sec. \ref{sec:conclusions}.

Hereafter, {Latin} indices run from $1$ to $3$, while {Greek} indices run
from $0$ to $3$, and the {signature} of the metric tensor is assumed to
be $(-,+,+,+)$. We also use the Einstein summation convention over
repeated indices and geometrised units in which the speed of light $c=1$
and the gravitational constant $G=1$. Appendix \ref{sec:appendixA} is
dedicated to the tedious but error-prone procedure needed to transform
from these units over to physical CGS units.

\section{Two-moment scheme for radiative transfer}
\label{sec:moment-scheme}

Before describing the equations of general-relativistic radiative
transfer (GRRT), we briefly summarizes the equations of ideal GRMHD that
describe the motion of the ordinary fluid (in the absence of radiation)
and that will need to be coupled to those describing the evolution of
the radiation fluid (see Sec. \ref{sec:coupling} for details on this
coupling).

\subsection{General-relativistic MHD}
\label{sec:GRMHD}

We recall that the ordinary fluid is described by the
conservation equations of mass and energy-momentum and by Faraday's
induction equation (with zero resistivity), \ie
\begin{align}
	\nabla_\mu (\rho u^\mu) &= 0 \,, \label{eq:grmhd1} \\
	\nabla_\mu T^{\mu\nu}_{\mathrm{fl}} &= 0 \,, \label{eq:grmhd2} \\
	\nabla_\mu \, ^{*}F^{\mu\nu} &= 0 \,, \label{eq:grmhd3} \,
\end{align}
with $\rho$ being the fluid rest-mass density, $u^\mu$ its four-velocity,
$T^{\mu\nu}_{\rm fl} $ the fluid energy-momentum tensor, which
includes contributions from the matter and the electromagnetic fields,
and the dual Faraday tensor $^{*}{F}^{\mu\nu}$.

In modern numerical codes, Eqs. \eqref{eq:grmhd1}--\eqref{eq:grmhd3} are
solved numerically after being cast in a conservative formulation in
order to assure numerical stability and convergence to the correct
solution in the presence of shocks \citep{Rezzolla_book:2013}. Overall,
they have the schematic form
\begin{align}
  \label{eq:evolutionfluid_0}
  \partial_t(\sqrt{\gamma}\, D) = \ldots\,,  \\
  \label{eq:evolutionfluid_1}
  \partial_t(\sqrt{\gamma}\, \tau) = \ldots\,,  \\
  \label{eq:evolutionfluid_2}
  \partial_t(\sqrt{\gamma}\, S_i) = \ldots\,, \\
  \label{eq:evolutionfluid_3}
  \partial_t(\sqrt{\gamma}\, B_i) = \ldots\,,
\end{align}
where we recall that $\gamma$ is the determinant of the spatial
three-metric, $D$ is the conserved rest-mass density, $\tau$ is the
rescaled total fluid energy density, $S_i$ the components of the
covariant three-momentum, and $B_i$ the components of the magnetic field,
all in the Eulerian frame [see, \eg, \citet{Porth2017} for details and
  the numerical methods normally employed to solve such equations].

In the presence of radiation, Eq. (\ref{eq:grmhd2}) has to be modified
since now $T^{\mu\nu}_{\mathrm{fl}}$ is no longer conserved, but rather
the total energy-momentum tensor is conserved, \ie
$T^{\mu\nu}=T^{\mu\nu}_{\mathrm{fl}}+T^{\mu\nu}_{\mathrm{rad}}$, where
$T^{\mu\nu}_{\mathrm{rad}}$ is the energy-momentum tensor of the
radiation fluid. It follows that
\begin{equation}
\nabla_\mu T^{\mu\nu}_{\rm fl} = -\nabla_\mu T^{\mu\nu}_{\rm rad} =: G^{\nu}\,
	\label{eq:conservation}\,,
\end{equation}
so that $G^\nu$ can be regarded as an external ``four-force''.

While we will provide an explicit expression for
$T^{\mu\nu}_{\mathrm{rad}}$ and $G^\nu$ in the next section, together
with the details on how to couple fluid and radiation in a numerical code
in Sec. \ref{sec:coupling}, we here only mention that when cast in a
conservative formulation, the evolution equations (\ref{eq:grmhd1}) and
(\ref{eq:grmhd3}) [or, equivalently, Eqs. \eqref{eq:evolutionfluid_0},
  \eqref{eq:evolutionfluid_3}] remain unaltered (the radiation fluid does
not alter the fluid's particle number and is not charged, thus does not backreact on the
background magnetic field), while (\ref{eq:grmhd2}) will need to be suitably
modified to account for the radiation contributions to the total energy
density and momentum.

Next, we describe our treatment of the radiation via a two-moment scheme,
which is widely used in radiation-hydrodynamics codes \citep[see,
  \eg][]{Roedig2012, Sadowski2013, McKinney2014, Oconnor2015,
  Foucart2015a, MelonFuksman2019}. In our implementation, we mostly
follow \citet{Foucart2015a}, which itself is based on the work of
\citet{Shibata2011} and \citet{Cardall2013b}.

\subsection{General-relativistic radiative transfer}
\label{sec:evolution}

Radiation in form of photons or neutrinos is described by their
distribution function $f(x^i, p^i, t)$, which depends on the spatial
coordinates $x^i$ and the particles momentum $p^i$ \citep[see, \eg][and
  references therein]{Rezzolla_book:2013}. This distribution function
changes in time according to the Boltzmann equation
\begin{equation}
\label{eq:boltzmann}
\left( p^{\mu} \frac{\partial}{\partial x^{\mu}} - \Gamma^{\mu}_{~\nu\rho}p^\nu p^\rho
	\frac{\partial}{\partial p^{\mu}} \right) f = \left( \frac{\partial
	f}{\partial \tau} \right)_{\rm{coll}}\,,
\end{equation}
where $\Gamma^{\mu}_{~\nu\rho}$ are the Christoffel symbols and $\tau$
the affine parameter of a radiation particle's trajectory. The right-hand
side includes the collisional processes such as emission, absorption and
scattering. Because Eq. (\ref{eq:boltzmann}) represents a 7D problem that
is, in general, too expensive to solve numerically, we adopt a formalism
in which the radiation field is described in terms of moments of the
distribution function and expressed in terms of projected, symmetric and
trace-free tensors \citep{Thorne1981}. In practice, we then employ an
approximation that involves the evolution of the lowest-two moments of
the distribution function \citep{Rezzolla1994,Shibata2011}. More
specifically, within a 3+1 decomposition of spacetime
\citep{Alcubierre:2008,Gourgoulhon2012}, one evolves the two moments of
the distribution function $f(x^i, p^i, t)$ that are thus defined as
\begin{align}
\label{eq:0moment_nu}
	J_{(\nu)} &:= \nu^3 \int_{4\pi} f(x^i, p^i, t)\, \rm{d}\Omega \,,\\
\label{eq:1moments_nu}
	H^{\mu}_{(\nu)} &:= \nu^3 \int_{4\pi} l^{\mu} f(x^i, p^i, t)\, \rm{d}\Omega\,.
\end{align}
where the integrals are taken in a frame comoving with the fluid (\ie the
``fluid frame''), $\nu$ (not to be confused with a tensor index) is the
radiation frequency and $d\Omega$ is the solid angle on a unit sphere in
momentum space and $l^{\mu}$ a unit normal four-vector orthogonal to the
fluid four-velocity $u^{\mu}$, \ie $l^{\mu}u_{\mu}=0$.

In practice, the quantities $J_{(\nu)}$ and $H^{\mu}_{(\nu)}$ represent the
frequency-dependent (hence the $(\nu)$ index) definitions of the
radiation energy density and of the radiation momentum density,
respectively. For simplicity, and to reduce computational costs, we here
limit ourselves to frequency-integrated moments, \ie
\begin{align}
\label{eq:0moment}
J &:= \int_0^\infty J_{(\nu)} \rm{d}\nu\,, \\
\label{eq:1moment}
	H^{\mu} &:= \int_0^\infty H^{\mu}_{(\nu)} \rm{d}\nu\,, 
\end{align}
thus to what is commonly referred to as "grey" approximation. Note that a
frequency-dependent scheme would require a discretization of the final
evolution equations in $\nu$ and thus increase the numerical cost by a
factor of $N$, where $N$ is the number of bins chosen for this
discretization. In addition, we define the second moment $L^{\mu\nu}$ as
\begin{equation}
\label{eq:Lmunu}
	L^{\mu\nu}: = \int_0^\infty \nu^3 d\nu \, \int_{4\pi} l^{\mu} l^\nu f
	\rm{d}\Omega\,.
\end{equation}
which represents the stress tensor of the radiation fluid.

Using these moments, it is possible to write the energy-momentum tensor
of the radiation as
\begin{equation}
\label{eq:Tradmunu}
	T^{\mu\nu}_{\rm{rad}} = J u^{\mu} u^\nu + H^{\mu} u^\nu + H^\nu u^{\mu} +
	L^{\mu\nu}\,.
\end{equation}
This tensor can also be written in the Eulerian frame as
\begin{equation}
\label{eq:Tradmunu2}
	T^{\mu\nu}_{\rm{rad}} = E n^{\mu} n^\nu + F^{\mu} n^\nu + F^\nu n^{\mu} +
	P^{\mu\nu}\,,
\end{equation}
where $n^{\mu}$ is a timelike unit four-vector normal to a hypersurface
when considering the $3+1$-decomposition of spacetime. The quantities
$E$, $F^{\mu}$ and $P^{\mu\nu}$ are respectively: the radiation energy
density, the radiation momentum density and the radiation pressure
tensor, all evaluated in in the Eulerian frame. Using the split of the
fluid four-velocity as $u^{\mu} = W(n^{\mu} + v^{\mu})$, where $W$ is the
Lorentz factor and $v^{\mu}$ the spatial four-velocity of the fluid in
the Eulerian frame, these quantities can be obtained from their
counterparts in the fluid frame via
\begin{align}
	E &= W^2 J + 2W v_{\mu} H^{\mu} + v_{\mu} v_\nu L^{\mu\nu}\,,
	\label{eq:fluid2euler_1} \\
	F_{\mu} &= W^2 v_{\mu} J + W(g_{\mu\nu} - n_{\mu} v_\nu) H^\nu
        + (g_{\mu\nu} - n_{\mu} v_\nu)v_\rho L^{\nu\rho}
        \nonumber \\
	& \quad
        + W\,v_{\mu} v_\nu H^\nu\,,
        \label{eq:fluid2euler_2} \\
	P_{\mu\nu} &= W^2 v_{\mu} v_\nu J + W(g_{\mu\rho} - n_{\mu}
	v_\rho)v_\nu H^\rho 
        + W (g_{\rho\nu} - n_\nu v_\rho)v_{\mu} H^\rho \nonumber \\
	&\quad + (g_{\mu\rho} - n_{\mu} v_\rho)(g_{\nu\lambda} - n_\nu
	v_\lambda)L^{\rho\lambda}\,,
        \label{eq:fluid2euler_3}
\end{align}
where $g_{\mu\nu}$ is the four-metric. Vice-versa, the fluid-frame
quantities can be obtained from the Eulerian ones via
\begin{align}
	J &= W^2 (E - 2 F^{\mu} v_{\mu} + P^{\mu\nu} v_{\mu}
	v_\nu)\,, \label{eq:euler2fluid_1} \\
	H^{\mu} &= W( E - F^\nu v_\nu)h^{\mu}_{~\rho} n^\rho + W
	h^{\mu}_{~\nu} F^\nu 
        - W h^{\mu}_{~\nu} v_\rho P^{\nu\rho}\,, \label{eq:euler2fluid_2} \\
	L^{\mu\nu} &= T^{\rho\lambda}_{\rm{rad}}\,h^{\mu}_{~\rho}
	h^\nu_{~\lambda}\,,
        \label{eq:euler2fluid_3}
\end{align}
where $h_{\mu\nu} := g_{\mu\nu} + u_{\mu} u_\nu$ is the projection tensor
orthogonal to the fluid four-velocity, \ie $h_{\mu\nu} u^{\mu}=0$. From
these equations it then follows that for any
fluid with $v_i = 0$, the following relations hold: $E=J$, $F^{\mu} =
H^{\mu}$ and $P^{\mu\nu} = L^{\mu\nu}$. We also note that $F^{\mu}$ and
$P^{\mu\nu}$ are purely spatial by construction, \ie $F^0 = P^{0\mu} =
P^{\mu 0}=0$.

The evolution equations for $E$ and $F_i$ in conservative form read
\begin{align}
  \partial_t (\sqrt{\gamma} E) &+ \partial_j(\sqrt{\gamma}(\alpha F^j -
  \beta^j E)) \nonumber \\
  &= \sqrt{\gamma}\,\alpha(P^{ij}K_{ij} - F^j\partial_j
  \textrm{ln}\,\alpha + G_0)\,, \label{eq:evolutionrad_1} \\
  \partial_t (\sqrt{\gamma}F_i) &+ \partial_j (\sqrt{\gamma}(\alpha P^j_{~i}
  - \beta^j F_i)) \nonumber \\
  &= \sqrt{\gamma}\,(F_j\partial_i \beta^j - E\partial_i
  \alpha + \frac{\alpha}{2}P^{jk}\partial_i \gamma_{jk} + \alpha
  G_i)\,,
  \label{eq:evolutionrad_2}
\end{align}

The right-hand sides of the
Eqs. \eqref{eq:evolutionrad_1}--\eqref{eq:evolutionrad_2} include -- in
addition to the ``geometric source terms'' such as the lapse $\alpha$,
the shift $\beta^i$, the three-metric $\gamma_{ij}$ and its determinant
$\gamma$, and the extrinsic curvature $K_{ij}$ -- the ``collisional
source terms''
\begin{equation}
\boldsymbol{G} = (G_0, G_i) = (-S^{\mu} n_{\mu}, S^{\mu} \gamma_{\mu i})
\,,
\end{equation}
where $S^{\mu}$ is written in terms of the fluid-frame quantities as
\begin{equation}
\label{eq:Smu}
S^{\mu} = \eta u^{\mu} - \kappa_a J u^{\mu} - \kappa H^{\mu}\,.
\end{equation}
Here, $\eta$ is the frequency-integrated emissivity, $\kappa_a$ the
frequency-averaged absorption opacity, and $\kappa := \kappa_a
+ \kappa_s$ is the total opacity, with $\kappa_s$ the frequency-averaged
scattering opacity. Formally, the definition of these coefficients
follows directly from integrating the Boltzmann equation over $\nu$ and
the corresponding expressions are therefore
\begin{align}
  \eta &\coloneqq \int_0^\infty \nu^3 \eta_{(\nu)} \rm{d}\nu \,, \label{eq:nu} \\
  \kappa_a &\coloneqq \frac{\int_0^\infty \nu^3 f \kappa_{a, (\nu)} \rm{d}\nu}
	{\int_0^\infty \nu^3 f \rm{d}\nu}\,, \label{eq:kappa_a} \\
	\kappa_s &\coloneqq \frac{\int_0^\infty \nu^3 f \kappa_{s, (\nu)} \rm{d}\nu}
	      {\int_0^\infty \nu^3 f \rm{d}\nu}\,. \label{eq:kappa_s}
\end{align}

These quantities essentially embody the coupling of the radiation fluid
with the matter fluid and are determined by the underlying microphysics,
\ie the constituents of the radiation fluid (neutrinos or photons) and
which interactions and reactions are taken into account. We detail our
choices for these parameters in Sec. \ref{sec:bondi}.

\subsection{Closure}
\label{sec:closure}

As it is common in moment-expansion approaches, given an expansion of the
distribution function at order $k$, the first $k$ evolution equations
involve the first $k + 1$ moments. Hence, when actually calculating a
solution, it is necessary to truncate the expansion and introduce a
\textit{``closure relation''},
namely, the $(k+1)$-th equation which specifies the value of the highest
moment used in terms of lower ones. This closure relation needs to be
derived on the basis of physical considerations and may differ from
problem to problem \citep{Thorne1981,Rezzolla1994}. In practice, what is
needed in our two-moment scheme is an explicit expression for the
radiation pressure tensor $P^{ij}$ in terms of lower-order moments, \ie
$E$ and $F^{j}$. Since it is possible to obtain explicit expressions for
$P^{ij}$ in the optically thin and optically thick (or ``diffusion'')
limits, $P_{\mathrm{thin}}^{ij}, P_{\mathrm{thick}}^{ij}$, we express the
closure relation as
\begin{equation}
  \label{eq:closure}
  P^{ij} = \frac{3\chi(\xi) - 1}{2}P_{\rm{thin}}^{ij} +
  \frac{3(1-\chi(\xi))}{2}P_{\rm{thick}}^{ij}\,,
\end{equation}
where $\chi(\xi)$ is the so-called \textit{closure-function} and $\xi$ is
the \textit{variable Eddington-factor} and is a measure of the degree of
anisotropy of the radiation fluid \citep{Rezzolla1994}. A possible
definition of $\xi$ is therefore
\begin{equation}
\label{eq:eddington}
\xi :=
\sqrt{\frac{h_{\mu\nu}H^{\mu} H^\nu}{J^2}} = \sqrt{\frac{H^{\mu}  
	H_{\mu}}{J^2}}\,,
\end{equation}
with the second equality holding because $H^{\mu} u_{\mu} = 0$. Note that
$\xi=1$ corresponds to the optically thin limit, while $\xi=0$ to the
optically thick one.

Another possible choice for the Eddington
factor [used \eg in \cite{MelonFuksman2019}] is instead
\begin{equation}
\label{eq:eddington_alt}
\xi := \sqrt{\frac{F^{\mu} F_{\mu}}{E^2}}.
\end{equation}
with $\xi = 0, 1$ still representing the two optical limits. We here
choose Eq. (\ref{eq:eddington}) over Eq. (\ref{eq:eddington_alt}),
although this means a substantially higher computational cost because of
the necessity of a root-finding method for computing $\xi$, which will be
detailed below. Nevertheless, Eq. (\ref{eq:eddington}) is the correct
choice since only this one is accurate in the optically thick limit
\citep[see][]{Shibata2011}.

For the {closure-function} $\chi$, we choose instead the so-called
\textit{Minerbo} closure (also referred to as the
\textit{maximum-entropy} closure) after
\citet{Minerbo1978} \footnote{This closure is often referred to as the
  M1 closure. However, this is a misnomer since the whole moment-scheme
  is usually called M1 scheme independent of the closure implemented.},
which is given by
\begin{equation}
  \label{eq:minerbo}
  \chi(\xi) = \frac{1}{3} + \xi^2 \frac{6 - 2\xi + 6\xi^2}{15}\,.
\end{equation}
We note that there are many other possible choices for the closure
function $\chi(\xi)$, \eg the often-used \textit{Levermore} closure
\citep{Levermore1984} and given by
\begin{equation}
\label{eq:levermore}
	\chi(\xi) = \frac{3 + 4\xi^2}{5+2\sqrt{4-3\xi^2}}\,.
\end{equation}
Hereafter, we will make use of Eq. (\ref{eq:minerbo}), but refer the
interested reader to \citet{Murchikova2017} for a comparison between
different closures.

We next calculate the radiation-pressure tensor in the two relevant
limits starting with the optically thin one ($\xi = 1$), recalling that
in this case $E^2 = F^{\mu} F_{\mu}$, so that we readily obtain
\begin{equation}
\label{eq:thin}
	P^{ij}_{\rm{thin}} = \frac{F^i F^j}{F^{\mu} F_{\mu}} E\,.
\end{equation}
For the thick limit ($\xi = 0$), on the other hand, we simply compute
$P^{ij}_{\rm{thick}}$ in terms of the thick-limit expressions for the
quantities in Eq. (\ref{eq:fluid2euler_3}), namely
\begin{align}
	L^{\mu\nu}_{\rm{thick}} &= \frac{J_{\rm{thick}}}{3}h^{\mu\nu}\,,
	\label{eq:thick_1} \\
	J_{\rm{thick}} &= \frac{3}{2W^2 + 1}[(2W^2 - 1)E -
	2W^2F^i v_i]\,,
	\label{eq:thick_2} \\
	(H_{i})_{\rm{thick}} &= \frac{F_i}{W} + \frac{W v_i}{2W^2 +
	1}[(4W^2+1) F^j v_j - 4W^2 E] \label{eq:thick_3}\,.
\end{align}

The difficulty with closing the system of Eqs.
(\ref{eq:evolutionrad_1})--(\ref{eq:evolutionrad_2}) lies in the
dependence of the Eddington-factor $\xi$ on $H^{\mu}$ and $J$, which, in
turn, depend on the unknown pressure tensor $P^{\mu\nu}$. It is
therefore necessary to obtain $\xi$ via a root-finding method as follows:
\begin{enumerate}
\item compute $P^{\mu\nu}_{\rm{thin}}$ and $P^{\mu\nu}_{\rm{thick}}$
  from $E$ and $F_i$ according to Eqs.
  (\ref{eq:thin})--(\ref{eq:thick_3}) and Eq. (\ref{eq:fluid2euler_3}).
\item[]
\item compute $P^{\mu\nu}$ according to Eq. (\ref{eq:closure}), where we set
  $\xi$ to the value of the previous timestep, and then use it in Eqs. 
  (\ref{eq:euler2fluid_1}) and (\ref{eq:euler2fluid_2}) to compute $J$ and 
  $H^{\mu}$.
\item[]
\item check if the function
  \begin{equation}
    \label{eq:rootfinding}
    f(\xi) = \frac{J^2 \xi - H^{\mu} H_{\mu}}{E^2}\,,
  \end{equation}
  is below a threshold value. If so, we have found the correct value for
  $\xi$. If not we adjust $\xi$ using a Newton-Raphson method, \ie
  \begin{equation}
    \label{eq:NewtonRaphson}
    \xi_{\rm{new}} = \xi - \frac{f(\xi)}{f'(\xi)}\,,
  \end{equation}
  where the derivative $f'(\xi)$ has to be computed via a
  finite-difference method, and then repeat the cycle. Alternatively one
  could find the root of $f(\xi)$ via Brent's method, which we find to be
  more robust, but also computationally more expensive.
\end{enumerate}

We should note that the choice of closing the system of evolution
equations with Eq. (\ref{eq:closure}) is computationally more expensive
than using the commonly used closure given by
\begin{equation}
  \label{eq:Lij}
  L^{ij}=L^{ij}_{\rm{thick}}\,.
\end{equation}
However, the assumptions behind the validity of the closure
\eqref{eq:Lij}, \ie isotropic radiation and $F^i F_i \ll E^2$, hold only
in the optically thick limit. Implementations with this choice of closure
can therefore model only those astrophysical scenarios where the optical
depth is high \citep[see, \eg][]{Roedig2012, Fragile2012}. Since we do
not wish to restrict to such conditions, our implementation with the
choice of Eq. (\ref{eq:closure}) allows to model both the thin and the
thick regimes.

\subsection{Computation of the fluxes}
\label{sec:fluxes}

When coupling \texttt{FRAC} with \texttt{BHAC} \citep{Porth2017}, which
solves the equations of GRMHD with finite-volume methods, it is simpler
to compute Eqs. (\ref{eq:evolutionrad_1}) and (\ref{eq:evolutionrad_2})
using the same finite-volume approach. To
accomplish this, we need, therefore, the interface-averaged ``fluxes''.
For second-order accuracy as the one employed here, these fluxes are
obtained by reconstructing the cell-averaged values of $E$ and $F_i$ to
the mid-points of the interfaces and then using an approximate Riemann
solver. Here, we use the minmod reconstruction and the HLL-Riemann solver
\citep[see][for an overview of these numerical
  methods]{Rezzolla_book:2013}, reconstructing $(E, F_i / E)$ rather than
$(E, F_i)$ as this then ensures causality. The characteristic speeds for
the Riemann solver depend on whether the fluid is optically thick or thin
and the limiting cases are again known exactly and given by
\citep{Shibata2011}
\begin{align}
\label{eq:characteristics}
	&\lambda_{\pm, \rm{thin}} = -\beta^i \pm \alpha \frac{F^i}{\sqrt{F_j F^j}}\,,
	\\
	&\lambda_{\pm, \rm{thick}} = \rm{min}\left(-\beta^i + p^i, \Lambda_\pm
	\right)\,,
\end{align}
with
\begin{align}
	&\Lambda_\pm = -\beta^i + \frac{2W^2 p^i\pm
	\sqrt{\alpha^2 \gamma^{ii}(2W^2 + 1) - 2(W p^i)^2}}{2W^2
	+ 1}\,,
\end{align}
where $p^i := \alpha v^i/W$ and $i$ denotes the direction in which the
characteristic speeds are evaluated. The final characteristic speeds are
then interpolated between the two regimes in the same manner as for the
radiation pressure tensor, \ie
\begin{equation}
\label{eq:interp_cs}
\lambda_\pm = \frac{3\chi(\xi) - 1}{2}\lambda_{\pm, \rm{thin}} +
	\frac{3(1-\chi(\xi))}{2}\lambda_{\pm, \rm{thick}}\,.
\end{equation}

The Eddington-factor $\xi$ at the cell interfaces is computed after the
reconstruction step, which is necessary since the fluxes depend on
$P^{\mu\nu}$. In Eq. (\ref{eq:interp_cs}) we can then simply use the same
$\xi$ and do not have to recompute it. We note that, because the
computation of $\xi$ via root-finding is the most expensive part of the
M1 scheme, we also tried other methods to reduce the computational
costs. A more efficient method is simply to interpolate $\xi$ from the 
surrounding cell-centres to the cell-faces (since the source terms also 
include $P^{\mu\nu}$, it
is necessary to compute $\xi$ in the cell centres anyway). Even more
efficient, albeit less accurate, would be to simply use the same value of
$\xi$ as computed for the cell-centres also at the cell interfaces. No
appreciable difference was found regarding the accuracy between all of
these methods, so that we adopted the latter, -- which is computationally
the least expensive -- as the default.

After obtaining the fluxes at the cell interface via use of an
approximate Riemann solver, we correct them to obtain the correct
asymptotic behavior also in the optically thick limit. As will be
discussed in Sec. (\ref{sec:implicit}), the collisional source terms, \ie
$G_0$ and $G_i$, become large for high optical depth $\kappa$, leading to
an inaccurate solution of the system
(\ref{eq:evolutionrad_1})--(\ref{eq:evolutionrad_2}) on a timescale of
$\mathcal{O}(1/\kappa \Delta x)$, where $\Delta x$ is the proper distance
(see below) between two adjacent grid cells \citep{Jin1996}. In essence,
this results into an incorrect diffusion rate of the radiation through
the fluid. In order to correct for this effect, we apply the same flux
corrections in the optically thick limit suggested by
\citet{Oconnor2015,Foucart2015a}, \ie
\begin{equation}
  \label{eq:fluxcorrection}
  \mathcal{F}_{E, \rm{corr}}^i = a\mathcal{F}_E^i + (1-a)\mathcal{F}_{E,
    \rm{asym}}^i\,,
\end{equation}
where $\mathcal{F}_E^i$ is the flux in Eq. (\ref{eq:evolutionrad_1}) in
$i$-th direction, $\mathcal{F}_{E, \rm{asym}}^i$ is the asymptotic flux
[see Eq. \eqref{eq:F_asym}], and $a$ is the weight function chosen as
\begin{equation}
\label{eq:a}
	a = \mathrm{tanh}\left(\frac{1}{\kappa_{\ell+1/2}\Delta x^i}\right)
\end{equation}
with $\kappa$ at the cell interface between the $\ell$-th and
$(\ell+1)$-th grid-cell approximated as $\kappa_{\ell+1/2} \approx
\sqrt{\kappa_\ell\, \kappa_{\ell+1}}$ and $\Delta x^i \coloneqq
\sqrt{\gamma_{ii}\left(\Delta x^i_{\mathrm{grid}}\right)^2}$ the proper
distance in $i$-th direction, with $\Delta x^i_{\mathrm{grid}}$ the
coordinate distance. The correct asymptotic flux in the optically thick
limit can be evaluated to be \citep{Thorne1981} 
\begin{align}
\label{eq:F_asym}
	\mathcal{F}_{E, \rm{asym}}^i &= \frac{4}{3}W^2\alpha v^i
	J_{\rm{thick}} - \beta^i E 
	-\frac{\alpha W}{3\kappa_{\ell+1/2}}(\gamma^{ij}+v^i
	v^j)\frac{dJ_{\rm{thick}}}{dx^j}\,.
\end{align}
and has to be computed on the cell interfaces. While $\kappa_{\ell+1/2}$ is
already a good approximation to its value on the interface, the other
quantities in the last term of Eq. (\ref{eq:F_asym}), \ie $\alpha$, $W$,
$\gamma^{ij}$ and $v^i$, are simply computed as the averages of
neighbouring cell-centered values. The total derivative of the energy
density in the fluid frame along the $i$-th direction is computed as
\begin{equation}
  \frac{dJ_{\rm{thick}}}{dx^i} = \frac{J_{\mathrm{thick}, \ell+1}-
    J_{\mathrm{thick}, \ell}}{\Delta x^i}\,.
\end{equation}
The quantities in the first two terms in Eq. (\ref{eq:F_asym}), on the other
hand,
are computed from their reconstructed left and right states via the
advection speed in $i$-th direction defined as
\begin{equation}
\label{eq:c_adv}
	c_{\rm{adv}} = -\beta^i + 4\alpha \frac{W^2}{2W^2 + 1}v^i\,.
\end{equation}
If $c_{\rm{adv}}$ is positive for both the left and right states, we
choose the left state; on the other hand, if it is negative for both the
left and right state, we choose the right state. In all other cases we
set all quantities in the first two terms in Eq. (\ref{eq:F_asym}) to
zero

Finally, to also ensure the correct behavior in the optically thick limit
for the fluxes of $F_i$, \ie for the quantities $\mathcal{F}_{F_i}^j$, we
choose to correct these fluxes as done in \citet{Audit2002}
\begin{equation}
  \mathcal{F}_{F_i, \rm{corr}}^j = b^2 \mathcal{F}_{F_i}^j +
  (1-b^2)\left(\frac{\mathcal{F}_{F_i, \ell+1}^j -
    \mathcal{F}_{F_i, \ell}^j}{2}\right)\,,
\end{equation}
\ie the flux at the cell interface is simply corrected with the average of
that flux in adjacent cell-centres weighted by a factor $b :=
1/(\kappa_{i+1/2}\Delta x^j)$. 

\subsection{Implicit treatment of stiff source terms}
\label{sec:implicit}

As mentioned above, the opacities $\kappa_a$ and $\kappa_s$ can become
very large for optically thick fluids. From Eq. (\ref{eq:Smu}) it is
evident that, under these conditions, also the collisional source terms
on the right-hand side of the evolution equations can become very large,
thus posing a major difficulty in solving these equations numerically. In
these regimes, the \emph{explicit} numerical solution of
Eqs. (\ref{eq:evolutionrad_1}) and (\ref{eq:evolutionrad_2}) requires a
prohibitively small timestep, making them ``stiff''. More specifically,
the timestep would have to be of order $\mathcal{O}(1/\kappa)$, thus
making the numerical evolution unfeasible. The situation is analogous to
that of resistive magnetohydrodynamics, where the timestep must decrease
with resistivity when employing explicit schemes \citep[see][for an
  implementation of similar methods for the case of general-relativistic
  resistive
  MHD]{Palenzuela:2008sf,Dionysopoulou:2012pp,Alic:2012,Ripperda2019}.

A solution to this limitation comes from the adoption of
mixed implicit and explicit methods.
In what follows we illustrate
the use of an implicit-explicit (IMEX) scheme that treats the advection
term and the geometric sources explicitly, while treating implicitly the
collisional source term. In such a scheme, a generic state vector
$\bm{U}$\footnote{We here denote with $\bm{U}$ the state vector relative to the
  radiation variables $E$ and $F_i$, while that of the fluid variables will
  be explicitly marked with a subscript as $\bm{U}_{\rm{fl}}$. From here on, we
  also write all state vectors and their corresponding flux
  and source vectors in boldface.}  is advanced
from timestep $n$ to the next timestep $n+1$ via $N+1$ 
intermediate steps denoted as
$\bm{U}^{(i)}$ ($i=0...N$) and given by \citep{Pareschi2010}
\begin{align}
	\bm{U}^{(i)} = \bm{U}^n + \Delta t \sum_{j<i} \tilde{a}_{ij}
	\bm{X}(\bm{U}^{(j)}) + \Delta
	t \sum_{j\leq i} a_{ij} \bm{M}(\bm{U}^{(j)})\,, \label{eq:IMEX_1} \\
	\bm{U}^{n+1} = \bm{U}^n + \Delta t \sum_{i=0}^{N} \tilde{w}_i
	\bm{X}(\bm{U}^{(i)}) + \Delta
	t \sum_{i=0}^{N} w_i \bm{M}(\bm{U}^{(i)}) \label{eq:IMEX_2}\,. 
\end{align}
The intermediate steps are computed from a combination of the
\emph{explicit} terms $\bm{X}$ and the \emph{implicit} terms $\bm{M}$, which are
weighted by the matrices $[\tilde{a}_{ij}]$ and $[a_{ij}]$,
respectively. These matrices are chosen so that the coefficients are zero
for $j\geq i$ in the explicit case and for $j>i$ in the implicit case. An
IMEX scheme with such matrices is referred to as the \textit{diagonally
  implicit Runge-Kutta} (DIRK) IMEX scheme. The matrix elements
$[\tilde{a}_{ij}]$ and $[a_{ij}]$, as well as the weights
$[\tilde{w}_{ij}]$ and $[w_{ij}]$, can be expressed conveniently via a
Butcher tableau and determine the specific type of the scheme and its
order. We here restrict ourselves to a second-order scheme with two
explicit and two implicit stages [called SS2(2,2,2) in the notation of
  \citet{Pareschi2010}]. In such a scheme the intermediate steps can be
written as
\begin{align}
	\bm{U}^{(0)} &= \bm{U}^n + \Delta t\, \gamma \bm{M}(\bm{U}^{(0)})\,, \\ 
	\bm{U}^{(1)} &= \bm{U}^n + \Delta t\, \bm{X}(\bm{U}^{(0)}) \nonumber \\
	&\quad + \Delta t\, \left((1-2\gamma)\bm{M}(\bm{U}^{(0)}) + \gamma
	\bm{M}(\bm{U}^{(1)})\right)\,, \\ 
	\bm{U}^{n+1} &= \bm{U}^n + \frac{1}{2}\Delta t\,
	\left(\bm{X}(\bm{U}^{(0)}) + \bm{X}(\bm{U}^{(1)})\right)
	\nonumber \\ 
	&\quad + \frac{1}{2}\Delta t\, \left(\bm{M}(\bm{U}^{(0)}\right) +
	\bm{M}(\bm{U}^{(1)}))\,,
\end{align}
where $\gamma = 1-1/\sqrt{2}$. Interestingly, it is possible to rewrite
the equations above in a way -- which corresponds to the one we have
actually implemented -- that avoids to store any intermediate explicit
term $\bm{X}$, namely, as
\begin{align}
	\bm{U}^{(0)} &= \left\{\bm{U}^n \right\} + \Delta t\, \gamma \bm{M}^{(0)}\,, \label{eq:SSP_1} \\ 
	\bm{U}^{(1)} &= \left\{\frac{3\gamma - 1}{\gamma}\bm{U}^n +
	\frac{1-2\gamma}{\gamma}\bm{U}^{(0)} + \Delta t\, \bm{X}^{(0)}\right\} 
	+ \Delta t\, \gamma \bm{M}^{(1)} \,,\label{eq:SSP_2} \\ 
	\bm{U}^{n+1} &= \left\{ \frac{1}{2}\left(\bm{U}^n + \bm{U}^{(1)} +
	\Delta t\,\bm{X}^{(1)}\right)
        +\Delta t\, \left(\gamma
	\bm{M}^{(0)}+\frac{1-\gamma}{2}\bm{M}^{(1)}\right)\right\}\,,
        \label{eq:SSP_3}
\end{align}
where we have also introduced a more convenient notation by writing
$\bm{M}/\bm{X}^{(i)}:= \bm{M}(\bm{U}^{(i)})/\bm{X}(\bm{U}^{(i)})$ and by using curly brackets to
highlight the explicit part of the equations. Such a notation helps to
see that Eqs. (\ref{eq:SSP_1}) and (\ref{eq:SSP_2}) are of the form
\begin{equation}
\label{eq:intermediate}
	\bm{U}^{(i)} = \bm{U}' + \mathrm{const.}\cdot \bm{M}^{(i)}\,,
\end{equation}
where $\bm{U}'$ is some intermediate state including all
contributions in the curly braces. Obviously and because of the implicit
nature of the scheme, the same state $\bm{U}^{(i)}$ appears both on the
left- and on the right-hand side of these equations. This implicit nature
of the equations for the intermediate states $\bm{U}^{(i)}$ generalises
also to higher-order IMEX schemes.

In general, an equation of the type \eqref{eq:intermediate} cannot be
solved analytically, which poses the biggest difficulty in using an IMEX
scheme and overall for the M1 scheme. In what follows we outline three
different strategies for solving these equations and start by detailing
how the quantities $\bm{X}^{(i)}$ and $\bm{M}^{(i)}$ are related to the evolution
equations (\ref{eq:evolutionrad_1}) and (\ref{eq:evolutionrad_2}).

As already mentioned, $\bm{X}$ includes all explicit terms, \ie
\begin{align}
\label{eq:X}
	&\bm{X}(\bm{U}^{(i)}) = \nonumber \\
	&\quad \left( \begin{array}{c}
		-\partial_k \mathcal{F}_E^k +
		\sqrt{\gamma}\alpha(P^{kl}K_{kl}-F^k\partial_k \rm{ln}\,\alpha) \\
		-\partial_k \mathcal{F}_{F_j}^k + \sqrt{\gamma}(F_k\partial_j
		\beta^k - E\partial_j \alpha + \frac{\alpha}{2}P^{kl}\partial_j
		\gamma_{kl})
	\end{array} \right)\,,
\end{align}
while $\bm{M}$ all the implicit terms, \ie
\begin{equation}
\label{eq:M}
	\bm{M}(\bm{U}^{(i)}) = \sqrt{\gamma}\alpha
	\left( \begin{array}{c}
		-S^{\mu} n_{\mu} \\
		S^{\mu} \gamma_{\mu j}
	\end{array} \right)= \sqrt{\gamma}\alpha
	\left( \begin{array}{c}
		G_0 \\
		G_i
	\end{array}
	\right)\,. 
\end{equation}

For each intermediate timestep $(i)$, there are three different ways of
performing the time update. The first and easiest method, which we refer
to as ``approximate method'' proceeds as follows:
\begin{itemize}
	\item[1.] compute $\bm{U}'$ in Eq. (\ref{eq:intermediate}) from
	  $\bm{U}^n$, $\bm{U}^{(j)}$ and $X^{(j)}$, where $j<i$.

	\item[2.] linearise the implicit term as
	  \begin{equation}
	    \label{eq:linearization}
		\bm{M}^{(i)} = \widehat{\bm{\mathcal{M}}}\, \bm{U}^{(i)} + \bm{b}\,,
	  \end{equation}
	  where the matrix $\widehat{\bm{\mathcal{M}}}$ and the column-vector
	  $b$ depend on the previous intermediate state $\bm{U}^{(i-1)}$ (or
	  $\bm{U}^n$ in the case of the zeroth step), which makes this
          linearisation only approximately true. Actually
	  $\widehat{\bm{\mathcal{M}}}$ and $\bm{b}$ would have to depend on the
	  current state $\bm{U}^{(i)}$ as well, in which case the
          linearisation would not be possible anymore. This approximation
          is justified under the assumption that the fluid four-velocity
          and the pressure tensor $P^{\mu\nu}$ do not change much during
          this intermediate timestep. The derivation of
	  $\widehat{\bm{\mathcal{M}}}$ and $\bm{b}$ is detailed below.

	\item[3.] solve Eq. (\ref{eq:intermediate}) via a matrix
          inversion
	  \begin{equation}
            \label{eq:inversion}
		  \bm{U}^{(i)} = \left(\widehat{\bm{\mathcal{I}}} - a_{ii}\Delta t\,
		  \widehat{\bm{\mathcal{M}}} \right)^{-1} \left(\bm{U}^{'} +
		  a_{ii}\Delta t\, \bm{b} \right)\,,
          \end{equation}
		where $\widehat{\bm{\mathcal{I}}}$ is the unit matrix.
\end{itemize}
In practice, this approximate method is the one used by
\citet{Foucart2015a}, although not within an IMEX scheme.

The second method, which is an improvement over the previous one and is
normally referred to as the ``fixed-point'' method, has been implemented
by \citet{Roedig2012, Fragile2014, MelonFuksman2019}, and is also used in the context of
resistive MHD by
\citet{Palenzuela:2008sf,Dionysopoulou:2012pp,Alic:2012,Ripperda2019}. In
such an approach, after step 3., the fluid four-velocity $u^{\mu}$ is
updated (see below for how the coupling between radiation and the fluid
is done) and from this, as well as from the new values for $E$ and $F_i$
the pressure tensor is recomputed. Going back to step 2., this procedure
is then iterated until the values of $\bm{U}^{(i)}$ for consecutive iterations
are below a given threshold value.

A third and final method, which we refer to as the ``root-finding''
method, consists in solving Eq. (\ref{eq:intermediate}) directly via a
root-finding procedure employing for example a four-dimensional
Newton-Raphson method
\citep{McKinney2014,Sadowski2013,MelonFuksman2019,Ripperda2019}.
Clearly, the fixed-point and the root-find methods are more complex and
computationally expensive, so that we here concentrate on results
obtained with the approximate method, postponing a detailed comparison
among the three methods to a future work.

Independently of which of the three methods discussed above is actually
used, it is necessary to write the collisional sources in terms of the
evolved variables $E$ and $F_{\mu}$, rather than in their counterparts in
the fluid frame $J$ and $H^{\mu}$. This can be done by using
Eqs. (\ref{eq:euler2fluid_1}) and (\ref{eq:euler2fluid_2}) in order to
replace $J$ and $H^{\mu}$ in Eq. (\ref{eq:Smu}). In this way, $S^{\mu}$
can then be written as
\begin{align}
\label{eq:Smu_2}
	S^{\mu} =& \left( \kappa_s W^2 u^{\mu} -\kappa W n^{\mu} \right) E 
        + \left( \kappa W v^i n^{\mu} - 2\kappa_sW^2 u^{\mu} v^i  \right) F_i \nonumber \\
	&+ \left( \eta u^{\mu} - \kappa_a W^2 u^{\mu} v_i v_j P^{ij} + \kappa W
	h^{\mu}_{~i}v_j P^{ij}  \right) + \kappa W F^{\mu}\,.
\end{align}
Note that it is not yet possible to write $S^{\mu}$ in the desired form
as $S^{\mu} = A^{\mu} E + B^{\mu i}F_i + C^{\mu}$, which is spoiled by
the last term in Eq. (\ref{eq:Smu_2}). This will only be possible after
contracting $S^{\mu}$ with $n_{\mu}$ and $\gamma_{\mu j}$, respectively,
as will be necessary for computing $\bm{M}^{(i)}$ according to
Eq. (\ref{eq:M}). After these contractions, the linearisation of
Eq. (\ref{eq:M}) follows as in Eq. (\ref{eq:linearization}) with
$\widehat{\bm{\mathcal{M}}}$ and $\bm{b}$ now given by
\begin{align}
  &\widehat{\bm{\mathcal{M}}} = \sqrt{\gamma}\alpha\, \times  \nonumber \\
  &\left(
  \arraycolsep=1.5pt
  \begin{array}{cccc}
    \kappa_s W^3 - \kappa W  & \zeta_0 W v^x & \zeta_0 W v^y & \zeta_0 W v^z     \\
    \zeta_1 \gamma_{jx} & \zeta_2 \gamma_{jx} W v^x  - \kappa W & \zeta_2 \gamma_{jx} W v^y  & \zeta_2 \gamma_{jx} W v^z  \\
    \zeta_1 \gamma_{jy} & \zeta_2 \gamma_{jy} W v^x                                         & \zeta_2 \gamma_{jy} W v^y  - \kappa W & \zeta_2 \gamma_{jy} W v^z  \\
    \zeta_1 \gamma_{jz} & \zeta_2 \gamma_{jz} W v^x                                         & \zeta_2 \gamma_{jz} W v^y             & \zeta_2 \gamma_{jz} W v^z - \kappa W  \\
  \end{array} \right)\,,
\end{align}
where we have introduced the shorthands
\begin{align}
\zeta_0 & := \kappa - 2W^2\kappa_s \,, \\
\zeta_1 & := \kappa_s W ^2u^j - \kappa W n^j \,, \\
\zeta_2 & := \kappa n^j - 2\kappa_s W u^j \,,
\end{align}
and
\begin{equation}
\label{eq:b}
	\bm{b} = \sqrt{\gamma}\alpha
   \left( \begin{array}{c}

           \eta W  + \kappa_s W  P^{\mu\nu}u_{\mu} u_\nu \\
	   (P^{\mu\nu}u_{\mu} (\kappa h^\alpha_{~\nu} - \kappa_a u_\nu u^\alpha) + \eta u^\alpha)\gamma_{\alpha x} \\
	   (P^{\mu\nu}u_{\mu} (\kappa h^\alpha_{~\nu} - \kappa_a u_\nu u^\alpha) + \eta u^\alpha)\gamma_{\alpha y} \\
           (P^{\mu\nu}u_{\mu} (\kappa h^\alpha_{~\nu} - \kappa_a u_\nu u^\alpha) + \eta u^\alpha)\gamma_{\alpha z} \\

   \end{array}\right) \,.
\end{equation}
As a consistency check, it is possible to verify that
\begin{equation}
\widehat{\bm{\mathcal{M}}}\cdot (E, F_i)^T + \bm{b} = \sqrt{\gamma} \alpha
(-S^{\mu} n_\mu , S^{\mu} \gamma_{\mu i})^T\,,
\end{equation}
with $S^{\mu}$ given by Eq. (\ref{eq:Smu}).

\subsection{Coupling between radiation and the ordinary fluid}
\label{sec:coupling}

At this point we can finally discuss the coupling between the radiation
and the ordinary (matter) fluid. As anticipated at the beginning of
Sec. \ref{sec:moment-scheme}, this is essentially done by ``adding'' to
the right-hand-sides of the conservative evolution equations of the fluid
[Eqs. \eqref{eq:evolutionfluid_1}--\eqref{eq:evolutionfluid_2}], the
components of the source-vector $\bm{M}_{\rm
  fl}=-\sqrt{\gamma}\alpha(G_0, G_i)^T$, \ie
\begin{align}
  \label{eq:evolutionfluid_1_rad}
  \partial_t(\sqrt{\gamma}\, \tau) = \ldots - \sqrt{\gamma}\alpha\, G_0\,,  \\
  \label{eq:evolutionfluid_2_rad}
  \partial_t(\sqrt{\gamma}\, S_i) = \ldots - \sqrt{\gamma}\alpha\,G_i\,.
\end{align}

As long as the ordinary fluid dominates, \ie the fluid variables have
values that are much larger than the corresponding radiation variables,
it is possible to simply treat these additional source terms explicitly,
as done in \citet{Roedig2012}. However, especially in regions of high
optical depth, these source terms may become much larger than the current
value of the fluid variables. In this case, the evolution equations for
the ordinary fluid, \ie Eqs. \eqref{eq:evolutionfluid_1_rad} and
\eqref{eq:evolutionfluid_2_rad}, become stiff and
have to be treated implicitly, exactly as we have already illustrated in
the case of the evolution equations for the radiation fluid, \ie
Eqs. \eqref{eq:evolutionrad_1}, \eqref{eq:evolutionrad_2}.

In practice, this is now rather straightforward and simply requires to
compute the source term of the fluid $\bm{M}_{\rm{fl}} =
(-\sqrt{\gamma}\alpha G_0,\, -\sqrt{\gamma}\alpha G_i)$
\begin{equation}
\label{eq:M_fluid}
	\bm{M}_{\rm{fl}} = - \bm{M} \,,
\end{equation}
and to add it to the state vector $\bm{U}'_{\rm{fl}}$, which is the
intermediate state in Eq. (\ref{eq:intermediate}), but this time for the
fluid variables.

Note that within this approach, the term $\bm{M}$ has not been computed
up to this point, because Eq. (\ref{eq:intermediate}) for the radiation
is solved via Eq. (\ref{eq:inversion}). However, with
$\widehat{\bm{\mathcal{M}}}$ and $\bm{b}$, which have already been
computed for this purpose, at every substep $(i)$ it is possible to
obtain easily $\bm{M}^{(i)}$ via Eq. (\ref{eq:linearization}) using the
newly computed $\bm{U}^{(i)}$ of the radiation. In this way,
Eq. (\ref{eq:intermediate}) is solved and the collisional source term is
automatically treated implicitly for the fluid as well, still assuming
that -- when using the approximate method -- $v_i$ and $P^{\mu\nu}$ are
approximately constant between the timestep $n$ and $n+1$. We note that
this assumption could be dropped for $P^{\mu\nu}$ when computing
$\bm{M}^{(i)}_{\rm{fl}}$, because one could easily recompute $P^{\mu\nu}$
from the newly computed $\bm{U}^{(i)}$. This choice, however, would break
energy-momentum conservation, because then $\bm{M}^{(i)}_{\rm{fl}} \neq
\bm{M}^{(i)}$. Clearly, when computing the collisional sources in
Eqs. (\ref{eq:evolutionfluid_1_rad}) and (\ref{eq:evolutionfluid_2_rad})
employing the fixed-point or the root-finding method, $\bm{M}^{(i)}$ (and
hence $\bm{M}^{(i)}_{\rm{fl}}$) would be known exactly after the
root-finding, so that hence energy-momentum conservation is automatically
satisfied in these latter two cases.

\texttt{FRAC} is implemented, so that it evolves the radiation variables for any
grid (note that Eqs. (\ref{eq:evolutionrad_1})-(\ref{eq:evolutionrad_2}) are
covariant, so that they work equally well with spherical, cylindrical and
cartesian grids.). In addition, it returns $\bm{\hat{\mathcal{M}}}$ and
$\bm{b}$ to the GRMHD code, from which it received the fluid (and metric) state
vector, so that the GRMHD code can easily add $\bm{M}^{(i)}_{\rm fl}$
to the fluid's source terms. Since this works grid-independent, $\texttt{FRAC}$
is particularly well suited for coupling to a code that implements block-based
adaptive mesh refinement (AMR) like \texttt{BHAC} (see also
below for AMR-related aspects).

In summary, the coupling between \texttt{FRAC} and any GRMHD code
only requires the following simple steps:
\begin{itemize}
	\item An initialisation between the GRMHD code and \texttt{FRAC} is
          necessary in order to set the grid and the spacetime in case it
          is fixed throughout the simulation (if it evolves dynamically,
          see the next point). The initialisation also includes the
	  addition of the four new radiation variables in the GRMHD code,
	  which while being evolved by \texttt{FRAC} are still stored in the
	  GRMHD code in order to employ its I/O and (possibly) mesh-refinement
	  routines.

	\item The GRMHD code has to pass at each intermediate timestep
          $(i)$ of the IMEX step the complete state vector of the fluid
          variables, $\bm{U}^{(i)}_{\rm{fl}}$, to \texttt{FRAC}, from
          which it computes $\eta$, $\kappa_a$ and $\kappa_s$. It then
          solves the implicit equation for $\bm{U}^{(i)}$ and returns
          $\bm{M}^{(i)}$. Optionally, the state of the metric can also be
          passed to \texttt{FRAC} at every timestep, if the spacetime
          does evolve dynamically.

	\item After computing all the explicit portions of the
          intermediate steps in the IMEX step within the GRMHD code,
	  $\bm{M}^{(i)}$ as received from \texttt{FRAC} can simply be
          subtracted with the correct pre-factor given by
          Eqs. (\ref{eq:SSP_1})--(\ref{eq:SSP_3}).

\end{itemize}
Obviously, these three steps assume that the GRMHD code already
implements an time integration like in
Eqs. (\ref{eq:SSP_1})--(\ref{eq:SSP_3}), which is fairly easy to do since
only the explicit portions of these equations need to be taken care of by
the GRMHD code. Consequently, only the pre-factors of the terms in the
curly brackets have to be implemented consistently.

Special attention has to be paid to the coupling with a GRMHD code that
employs AMR techniques, as is the case for
\texttt{BHAC}. In general, \texttt{FRAC} works independently of the
underlying grid structure. However, as is the case for the fluid
variables of any GRMHD code employing AMR techniques, also the fluxes of
the radiation variables have to be corrected at the interfaces between
coarse and fine grid cells. Fortunately, this operation can be handled in
perfect analogy with what is done for the fluid variables. More
specifically -- assuming that the GRMHD code has already developed all
the necessary operators for the prolongation and restriction procedures
needed when refining or coarsening the grid, respectively --
\texttt{FRAC} only needs to provide the GRMHD code with the radiation
fluxes $\mathcal{F}_E^j$ and $\mathcal{F}_{F_i}^j$ at the grid locations
where the fluid fluxes are computed before the AMR step. The prolongation
and restriction operators will then treat the radiation fluxes exactly as
the other fluid fluxes, providing AMR values for all variables.

\section{Special-relativistic tests}
\label{sec:tests}

We next present a number of standard-tests that have been performed in
order to verify our implementation of the M1 scheme. We will start by
considering below tests carried out in special relativity, which serve as
a preparation for the tests carried out in section \ref{sec:grtests},
which are instead performed in a curved but fixed spacetime. We also note
that all tests presented in sections
\ref{sec:beamtest}--\ref{sec:scattering} have been executed without the
coupling to a GRMHD code and thus probe \texttt{FRAC} as a stand-alone
code for dynamically evolving radiation. The tests presented in section
\ref{sec:fluid-coupling}, on the other hand, do refer to a situation in
which radiation is coupled to an ordinary fluid evolved with
\texttt{BHAC}.

\subsection{Straight-beam tests}
\label{sec:beamtest}
%
\begin{figure}
\begin{center}
  \includegraphics[width=0.99\columnwidth]{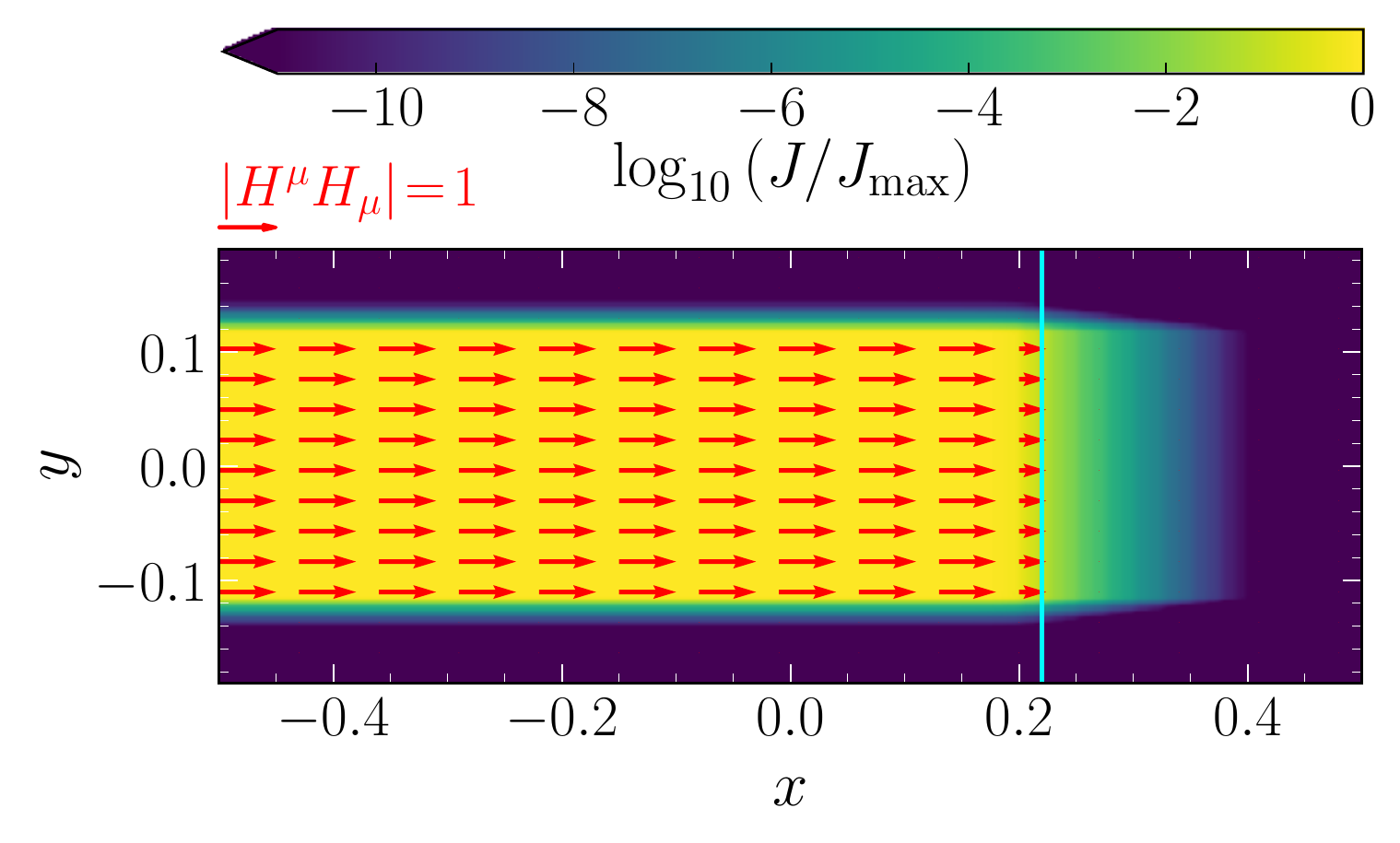}
  \caption{Propagation of a radiation beam that is injected from the left
    boundary in the domain. The beam energy density is colourcoded, while
    the fluxes are shown as a vector field with red arrows. Note the
    small amount of diffusion ahead of the beam edge, indicated with a
    cyan line.}
  \label{fig:beam_straight}
\end{center}
\end{figure}%
\begin{figure}
\begin{center}
  \includegraphics[width=0.99\columnwidth]{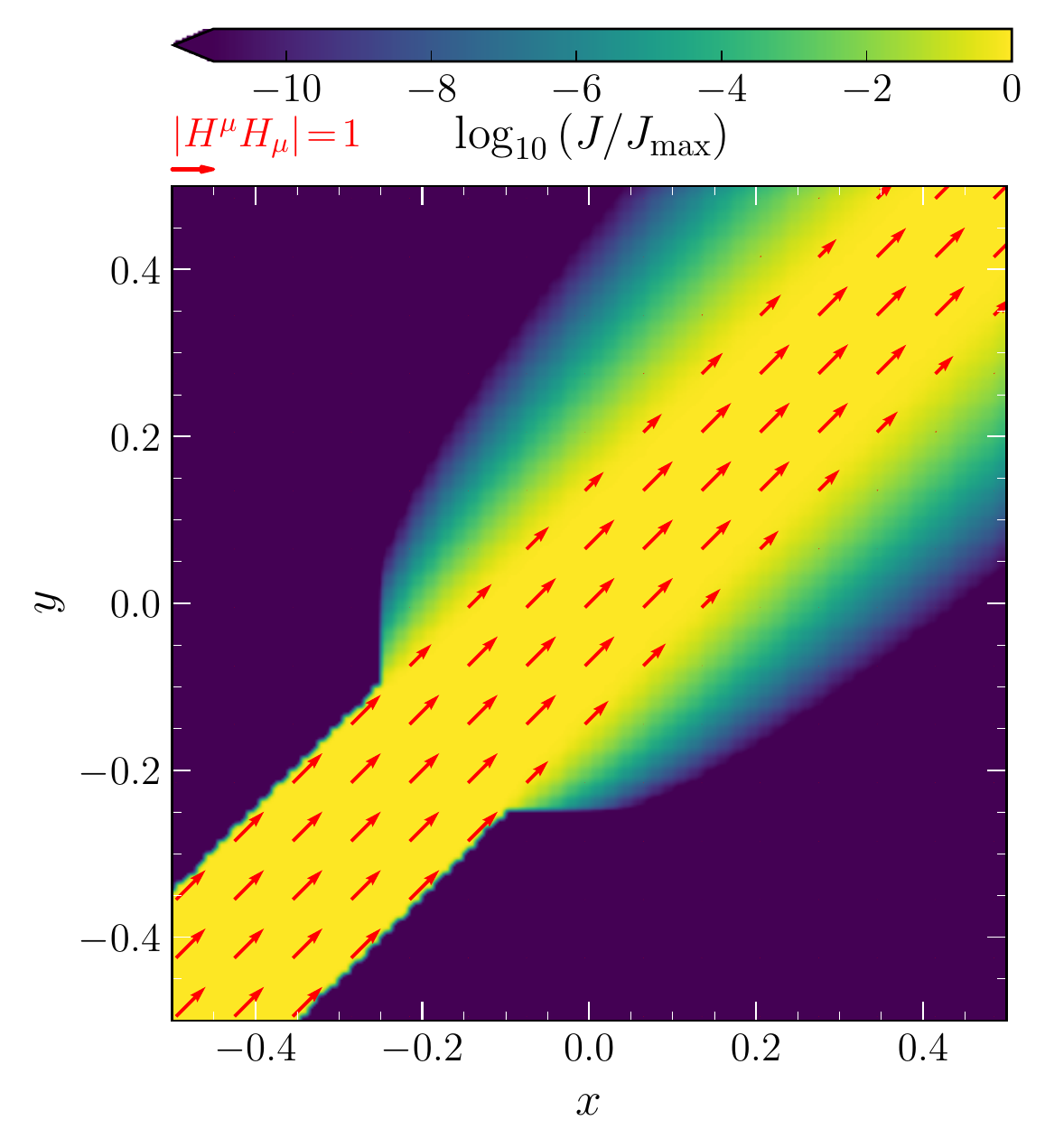}
	\caption{Same as Fig. \ref{fig:beam_straight}, but for a beam
          injected diagonally from the bottom left corner. The region defined
	  by$x<-0.25$ and $y<-0.25$ is enforced via a boundary condition
	  ensuring a continuous inflow of radiation.}
  \label{fig:beam_diag}
\end{center}
\end{figure}%
\begin{figure}
\begin{center}
  \includegraphics[width=0.99\columnwidth]{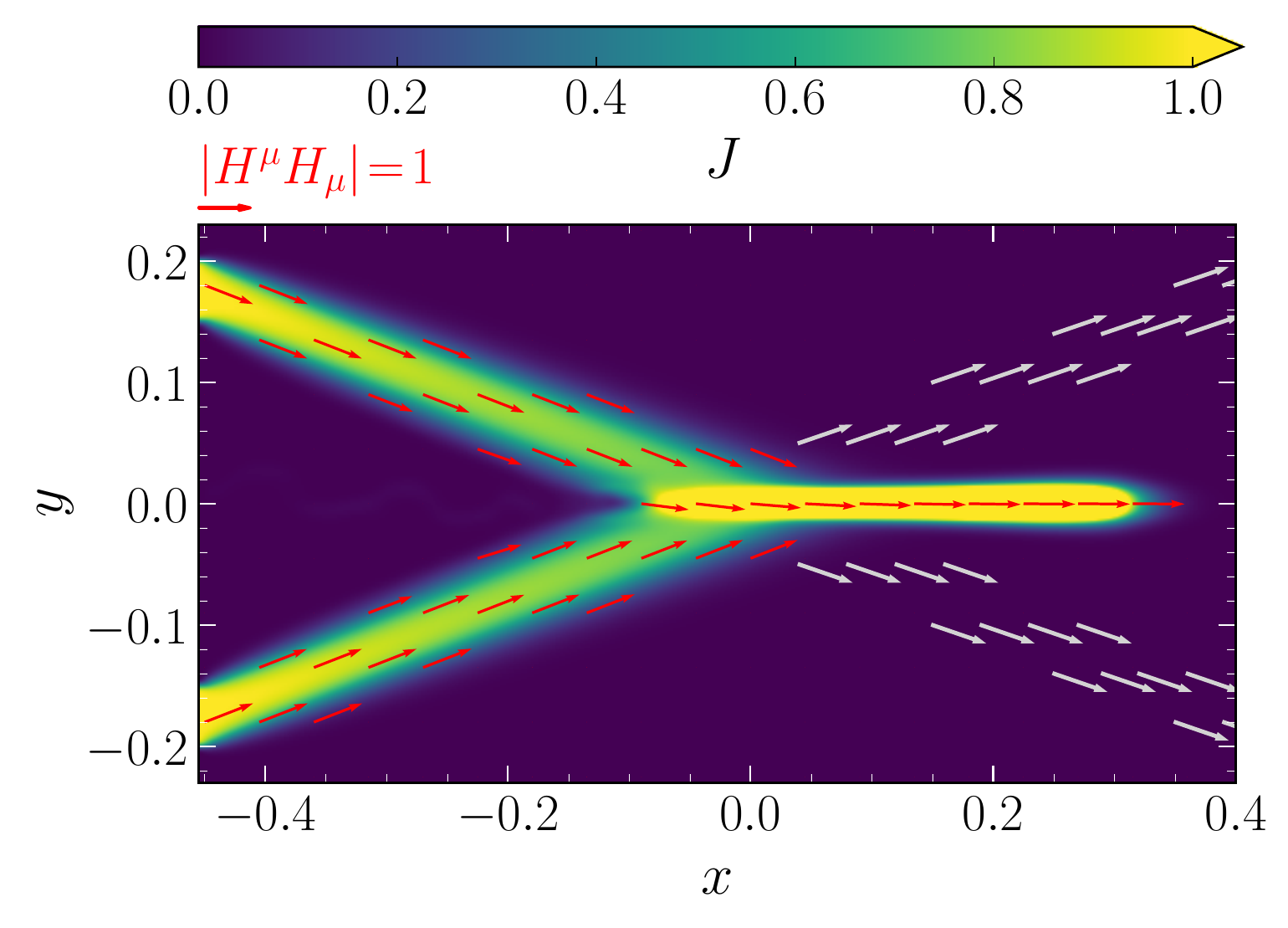}
  \caption{Same as Fig. \ref{fig:beam_straight}, but for two beams
    injected from the top left and bottom left corners. Note that instead
    of crossing each other at $y=0$ (\cf white arrows), the two beams
    merge into a single one, whose direction of propagation is the
    average of the original beams. To illustrate the failure of the M1
    scheme in this problem, we adopted a linear colourcode as opposed to the
    logarithmic one in Figs. \ref{fig:beam_straight} and
    \ref{fig:beam_diag}.}
  \label{fig:beam_cross}
\end{center}
\end{figure}%

As a first test we consider the propagation along a coordinate axis of a
straight beam of radiation in flat spacetime and in vacuum. To this scope
we consider a two-dimensional setup in Cartesian coordinates with domain
$-0.5 < x < 0.5,\, -0.2 < y < 0.2$, which we cover with $100 \times 40$ cells.
The radiation fluid
is initialised having a energy density given by
\begin{equation}
  \label{eq:straightbeam}
  J=\begin{cases}
  1\,,& ~ x<-0.4 ~ \textrm{and} ~~ |y| < 0.12 \\
  10^{-15}\,,& ~ \textrm{otherwise}
  \end{cases}
\end{equation}
and the fluxes as $H_x = J$, $H_y=H_z=10^{-15}$ (we recall that in all
tests with a static background fluid $J= E$ and $H_i = F_i$). The initial
data in \eqref{eq:straightbeam} is meant to simulate a beam that is shot
into the domain from the left
boundary. As expected in the absence of any sources such as gravity or
collisions, the beam of radiation should propagate parallel to the
$x$-axis from left to right at the speed of light. Figure
\ref{fig:beam_straight} shows that this is indeed the case and reports
the radiation energy density in a colorcode scale and with the red arrows
indicating the direction of motion of the radiation fluid. Note that the
leading edge of the radiation beam suffers from a certain amount of
diffusion in the longitudinal direction, which is an inevitable
consequence of the use of a grid-based code and disappears with
resolution.

A more demanding scenario is that of a straight beam that does not move
in a direction parallel to the coordinate axes, but at a certain angle
(45 degrees here). Figure \ref{fig:beam_diag} shows such a configuration
moving diagonally through the domain of size $-0.5 < (x,y) < 0.5$ with $100
\times 100$ cells. Here
we apply a boundary-condition that freezes the initial configuration for
$x<-0.25$ and $y<-0.25$. As expected diffusion is now more prominent and
present also in the direction orthogonal to the direction of
propagation. We also note that the setup for these tests is chosen so
that $H_{\mu} H^{\mu} = J^2$. In this limit, and because $v_i=0$, the
variable Eddington-factor $\xi$ should always be $1$. We monitor $\xi$
during the simulations and verify that our closure does indeed yield the
correct result in this optically thin limit with a precision that is set by the
threshold that we choose for the root-finding described in Sec.
\ref{sec:closure}. 

Finally, in Fig. \ref{fig:beam_cross} we simulate the case of two beams
that meet each other along the $y=0$ coordinate direction. Assuming the
radiation to be photons or neutrinos of the same flavor, one would expect
the two beams to cross without interacting and thus to continue on
straight paths (the expected direction of propagation is indicated with
white arrows). However, it is known that the moment-scheme, which treats
the radiation like a fluid, does not perform well in this scenario
\citep{Fragile2014,McKinney2014,Foucart2015a,RiveraPaleo2019}. Indeed,
Fig. \ref{fig:beam_cross} shows that the two beams merge into a single
beam of increased energy density that propagates along the $y$-direction,
namely in the direction resulting from the average of the original
propagation directions. This incorrect behavior can be understood when
considering that the moments are integrals of the distribution function
over the momentum space. While the distribution function stores the
information about all possible directions of propagation, its moments
lose this information as a result of the integration and thus provide
only a single averaged direction of propagation. As a result, the momenta
in opposite directions cancel and the information on the original
momentum distribution is lost. In principle, such information could be
recovered through the use of higher moments but is inevitably lost here,
where the second moment is only approximated analytically.

We note that while the moment scheme performs well for divergent
radiation in the optically thin limit (and rather generally for the
optically thick limit), the pathologies described here for the
crossing-beam problem would lead to a rather unphysical behavior if the
M1 scheme is applied to a realistic simulation of a merger of binary
neutron stars and when a black hole is formed as a result of the collapse
of the post-merger object. In this scenario, in fact, in which the black
hole is surrounded by a torus emitting neutrinos in all directions, the
solution of the M1 scheme along the polar axis of the black hole would be
incorrect, possibly leading to an overestimation of the radiation energy
density in the system's polar region \citep{Foucart2018}. In order to
solve this problem, different methods for treating radiative transport
are required and an alternative to the commonly adopted Monte Carlo
method \citep{Foucart2017,Miller2019a} will be presented elsewhere.

\subsection{Radiation wave in free-streeming regime}
\label{sec:radiationwave}
%
\begin{figure}
\begin{center}
  \includegraphics[width=0.995\columnwidth]{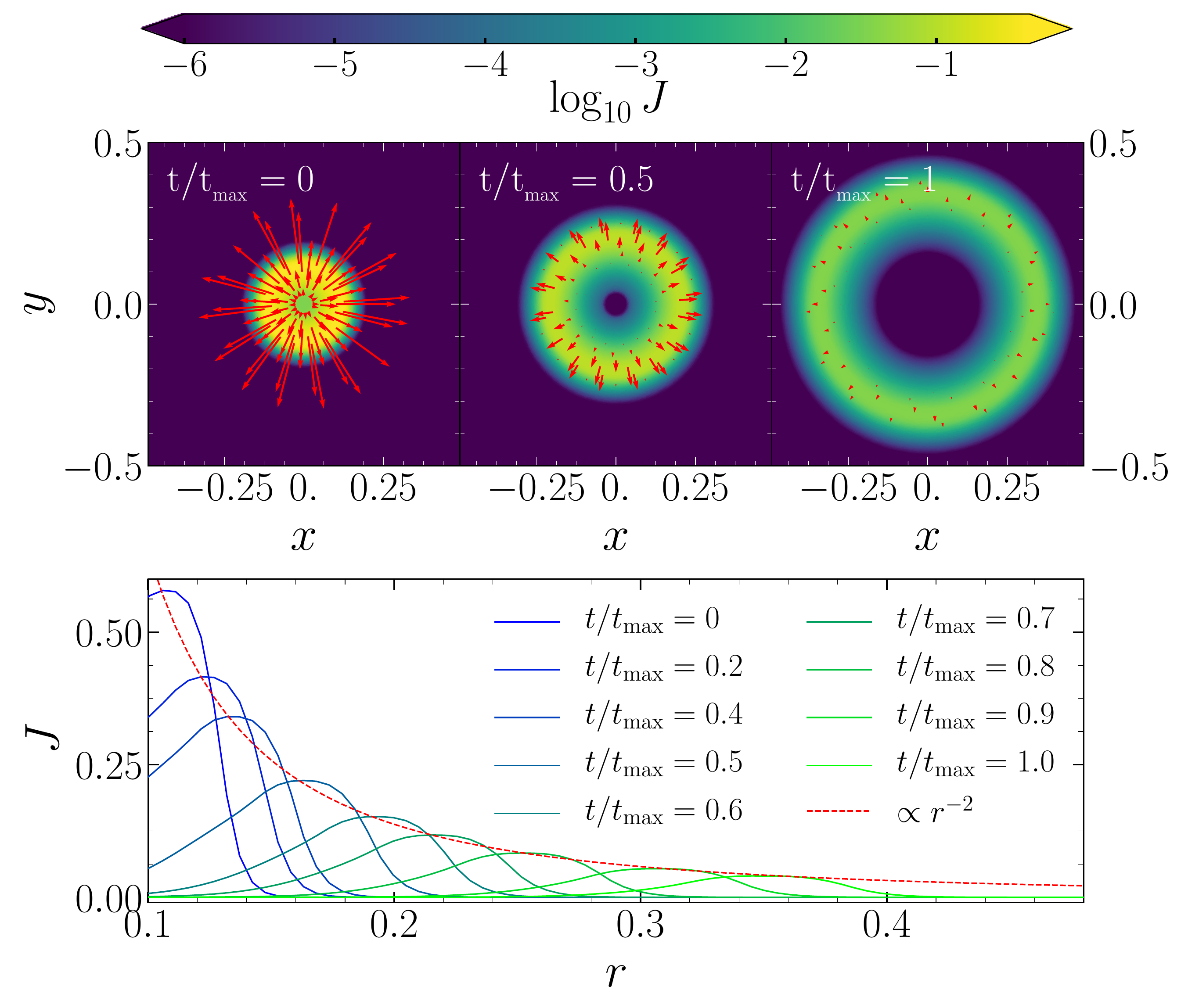}
  \caption{\textit{Top:} Same as Fig. \ref{fig:beam_straight} but for a
    radiation wave emitted radially from the origin of
    coordinates. \textit{Bottom:} Profiles of the energy density at
    different times (blue to green solid lines) and the comparison to a
    functional dependency $\propto 1/r^2$ (red-dashed line).}
  \label{fig:freewave}
\end{center}
\end{figure}%
The above beam tests work particularly well on a Cartesian grid. Our
generic implementation of the moment-scheme together with \texttt{BHAC}'s
ability to also handle non-Cartesian coordinates, allows us to also
perform simulations on spherical grids. To test this capability, we
consider a wave of radiation that freely propagates over the grid. We do
so by initialising a constant energy density for the radiation in a
circular region around the origin of a two-dimensional polar grid. Within
this region, we set $H_r = J$, while outside of it the energy density and
the fluxes are set to zero. As in the previous tests, we assume the
background to be vacuum via setting $\eta = \kappa_a = \kappa_s = 0$
throughout the simulation, so that no interaction with the fluid can take
place.

As can be seen from the top panels in Fig. \ref{fig:freewave}, the
radiation propagates in a ring-like structure over the grid. The initial
energy density spreads over this ring, whose radius increases in
time. Conservation of energy dictates, therefore, that the maximum of the
energy density decreases as the wave propagates. In our spherical grid,
this decrease is expected to happen at a rate $\propto 1/r^2$, which we
can verify by plotting a one-dimensional cut through the ring at
different time; this is shown in the bottom panel of
Fig. \ref{fig:freewave}. Taking the maxima of these profiles, we can then
fit a function of the form $f(r) \propto 1/r^2$ to the data, which is
shown as a red-dashed line. Clearly, we find good agreement between the
decrease of the energy density's maximum value and this functional
dependence, with relative deviations that are of $\Delta J/J = 4.7\times
10^{-2}$ at most.

\subsection{Shadow test}
\label{sec:shadowtest}
%
\begin{figure}
\begin{center}
  \includegraphics[width=0.995\columnwidth]{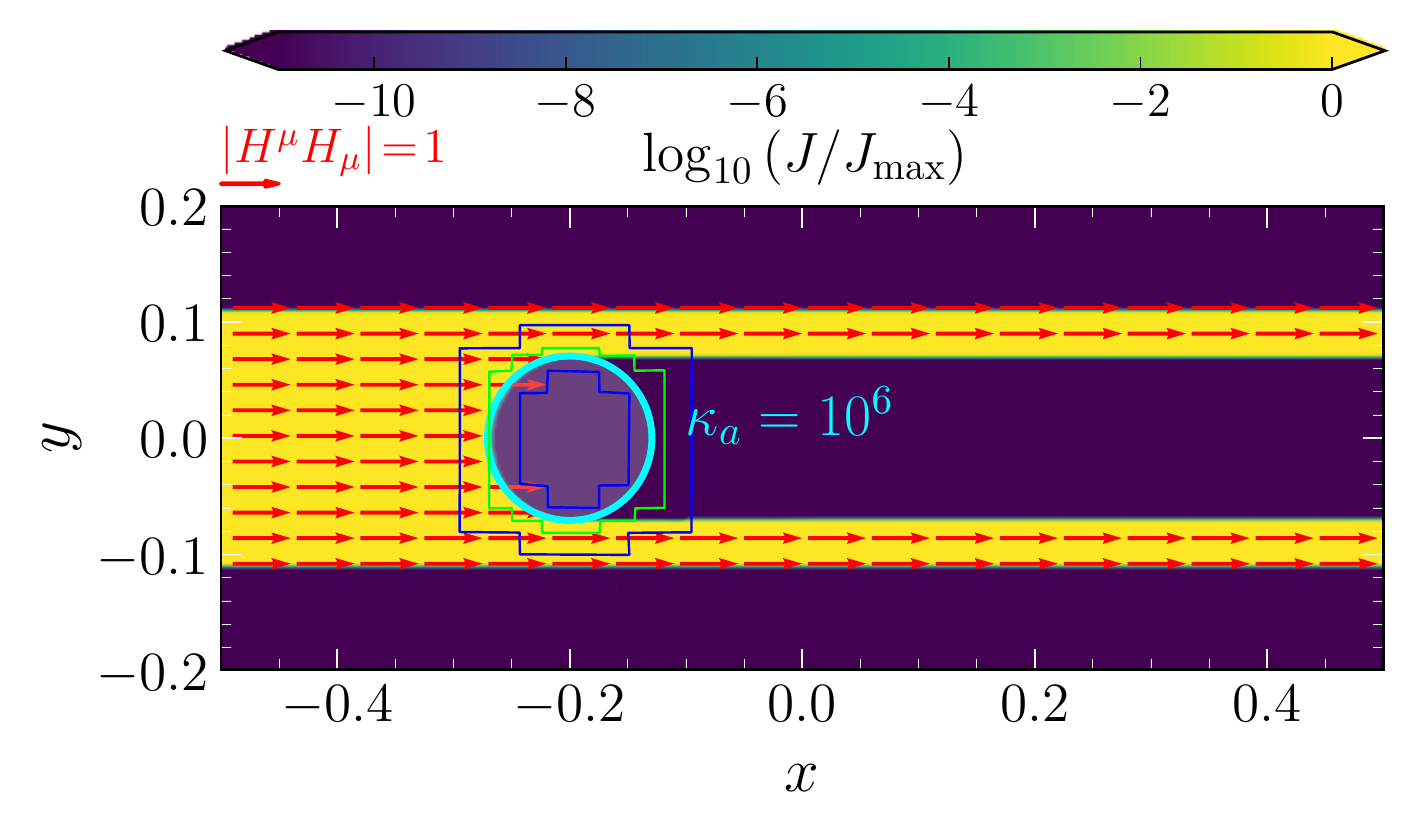}
  \caption{Same as Fig. \ref{fig:beam_straight} but when the beam hits an
    optically thick sphere (region within the cyan circle) yielding a
    shadow downstream of the sphere. Blue and green contours show the
    boundaries of the first and second mesh-refinement levels,
    respectively.}
  \label{fig:shadow}
\end{center}
\end{figure}%

Next, we simulate the interaction between radiation and fluid by placing
a dense sphere in the beam's path. However, rather than placing a static
and rigid spherical fluid configuration whose evolution we are not
interested in, we simply fix the absorption opacity in Eq. (\ref{eq:Smu})
in the region where we want the radiation to be absorbed. In particular,
we set $\kappa_a=10^6$ within a sphere of radius $r=0.07$ and origin $(x,
y) = (-0.2, 0)$ on the same domain as chosen for the beam in
Fig. \ref{fig:beam_straight}, but this time with two additional
refinement levels in order to better resolve the sphere. In
Fig. \ref{fig:shadow} the outline of this sphere is shown with a cyan
circle, while the boxes of different colours represent the adaptive mesh
structure adopted in this test. The same figure also shows how the beam
is obstructed by this optically thick sphere. This results in a shadow
behind the sphere and the splitting of the original beam into two beams
on the top and the bottom of the sphere, which themselves remain well
collimated and with little diffusion in the transverse direction. Where
the beam meets the high-opacity circle, a small amount of radiation is
expected to diffuse inside the region of absorption due to the finite
grid-resolution and the finite value of $\kappa_a$. Given the mesh
refinement and the relatively high value of $\kappa_a=10^6$ we only
measure a negligible amount of radiation diffusing inside the sphere.

It is useful to remark that the value of the opacity $\kappa_a$ is about
six orders of magnitude higher than that of the radiation variables. As a
result, the set of evolution equations become very stiff and we are able
to obtain a stable solution only thanks to the use of the IMEX
scheme. Indeed, we have verified that without decreasing the timestep to
prohibitively small values, an explicit time integration would yield a
stable solution only for $\kappa_a\sim 1$.

\subsection{Radiating sphere}
\label{sec:radsphere}
%
\begin{figure}
\begin{center}
  \includegraphics[width=0.9\columnwidth]{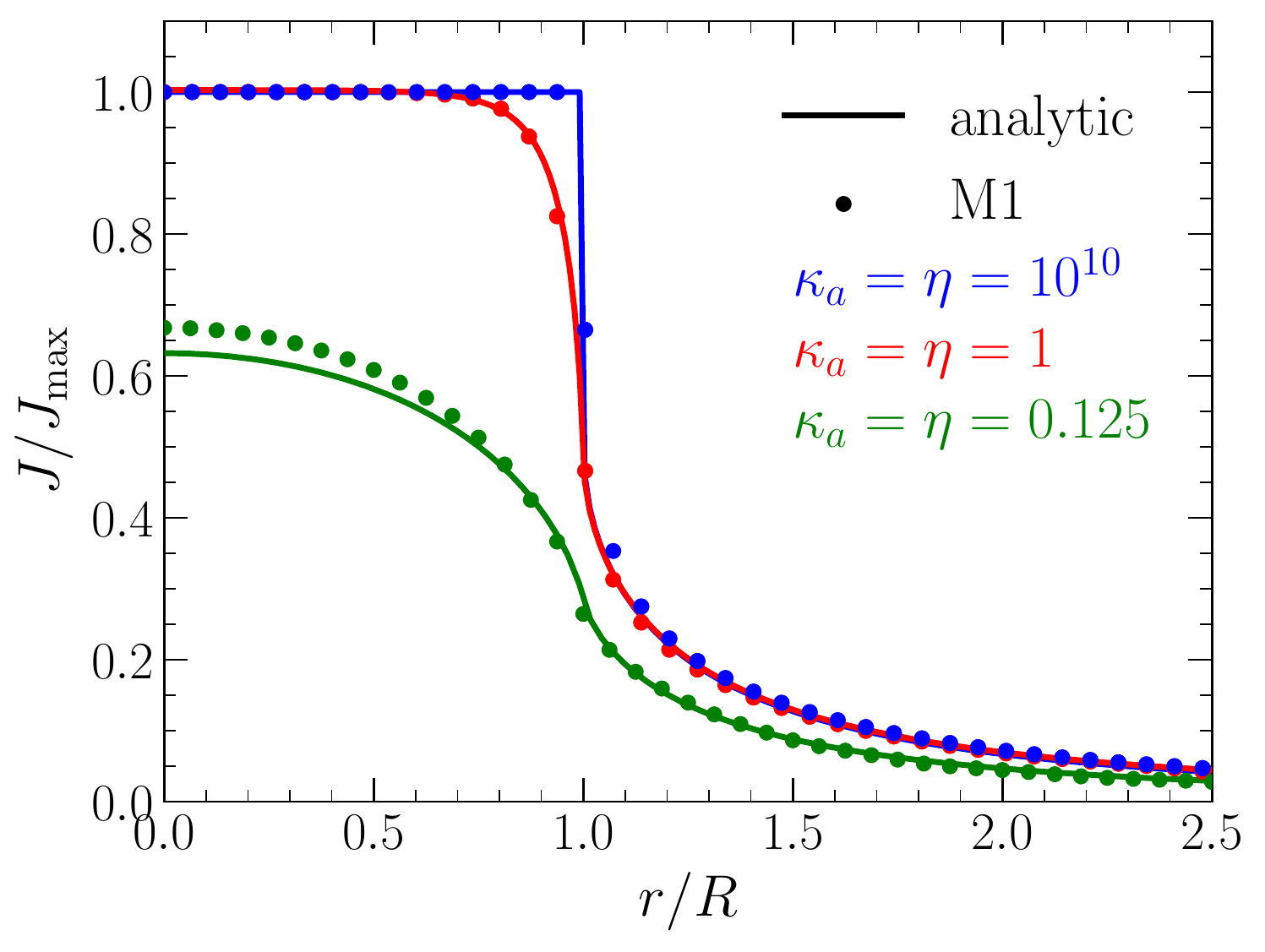}
  \caption{Equilibrium radiation energy density for a radiative
    sphere. The solid lines show the analytic solution according to
    Eq. (\ref{eq:radsph}), while the filled circles show the numerical
    results, for small (green), medium (red) and high (blue) values of
    $\kappa_a$ and $\eta$.}
  \label{fig:rad_sphere}
\end{center}
\end{figure}%
\begin{figure*}
\begin{center}
  \includegraphics[width=0.99\textwidth]{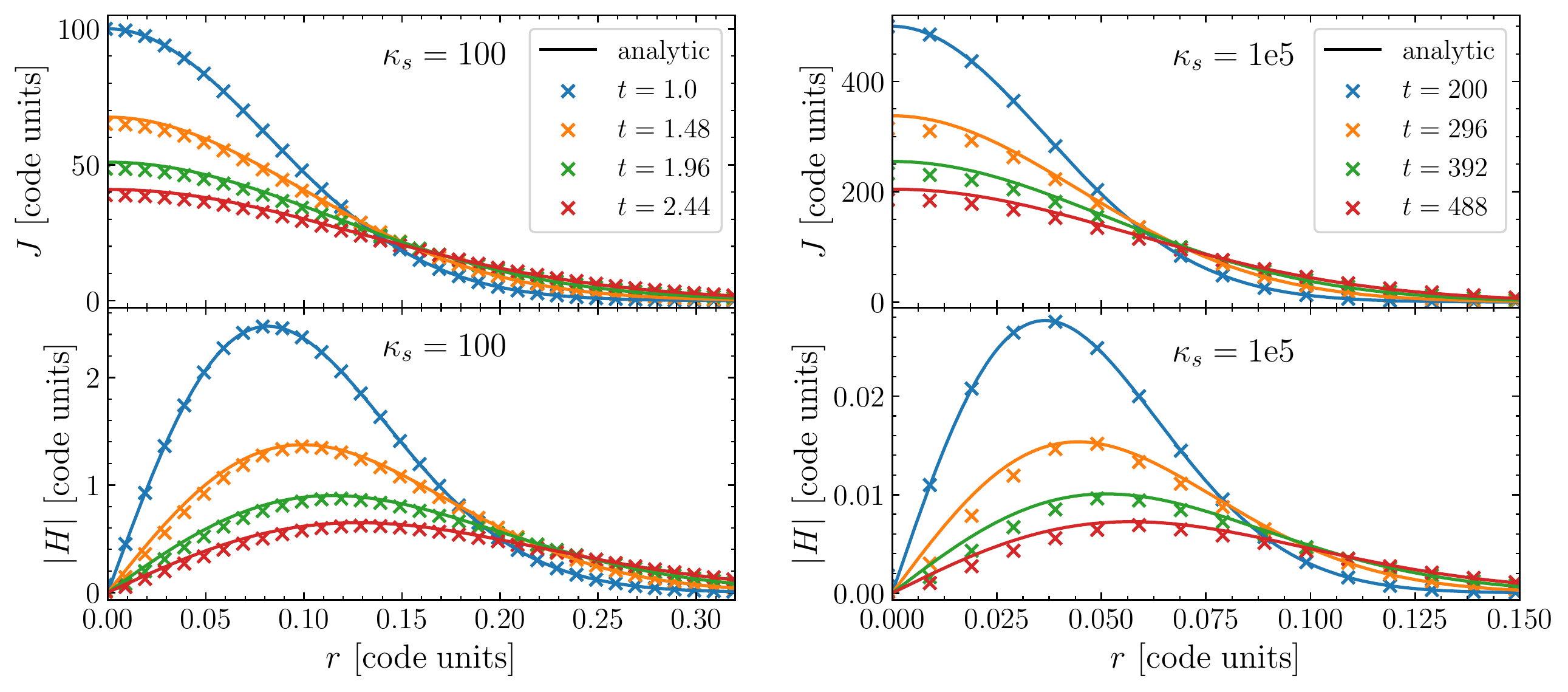}
	\caption{Radiation energy density and momentum density for the
          diffusion-wave test with $\kappa_s=100$ (left) and
          $\kappa_s=10^5$ (right; note that the linear scale is smaller
          than in the left panel). Crosses mark the numerical results at
          different times and solid lines the analytical solution
          according to Eqs. (\ref{eq:gauss1}) and (\ref{eq:gauss2}).}
  \label{fig:scattering}
\end{center}
\end{figure*}
While the assessment of the correctness of the previous tests was
essentially qualitative and based on how the beams of radiation should
propagate, we now perform a test, for which an analytic solution is
known, thus provide a more stringent and quantitative assessment. In
particular, we consider the homogeneous-sphere test first proposed by
\citet{Smit:97}. We again perform the test in vacuum and artificially
introduce a fluid by adjusting the opacities and the emissivity. More
specifically, we set $\kappa_s=0$ everywhere and
$\kappa_a=\eta=\textrm{const.}$ within a sphere of radius $R$ and
$\kappa_a=\eta=0$ everywhere else. This setup can be thought of as
representing a sphere of radiation with constant energy density and that
radiates in equilibrium. A possible physical interpretation could
therefore be an isolated and radiating hot neutron star. While a neutron
star does not have a constant rest-mass density, the sharp drop of
$\kappa_a$ and $\eta$ to zero at the surface provides a rather realistic
description of the extreme transitions expected near the stellar surface,
where the density drops to zero over a very narrow region.

The distribution function for this model is known analytically in terms
of the radius $r$ and of the azimuthal angle $\theta$. After setting
$\mu:=\rm{cos}\,\theta$, it reads
\begin{equation}
  f(r,\mu) = b(1-e^{-\kappa_a s(r,\mu)})\,,
\end{equation}
where $b$ is a constant that can be freely specified (see below) and 
\begin{equation}
  s:=\begin{cases}
  r\mu + Rg(r,\mu)\,, & r<R ~ \land ~ -1<\mu<1 \\
  2Rg(r,\mu)\,, & r\geq R ~ \land ~ \sqrt{1-R^2/r^2} <
  \mu < 1
  \end{cases}
\end{equation}
and
\begin{equation}
  g(r,\mu) := \sqrt{ 1-\frac{r^2}{R^2}(1-\mu^2)}\,.
\end{equation}
The zeroth moment $J = E$ can then be obtained via integration of the
distribution function as [\cf Eq. \eqref{eq:0moment}] 
\begin{equation}
	J(r) = \frac{1}{2} \int_{-1}^1 d\mu\, f(r,\mu) \,.
        \label{eq:radsph}
\end{equation}

In Fig. \ref{fig:rad_sphere} we show the solution of
Eq. (\ref{eq:radsph}) for a small, medium and a high value of the
emissivity and absorption opacity, \ie for $\kappa_a=\eta=0.125$ (green),
$\kappa_a=\eta=1$ (red) and $\kappa_a=\eta=10^{10}$ (blue),
respectively. Solid lines of different colours show the analytic
solutions Eq. (\ref{eq:radsph}), while the filled circles the
corresponding numerical results; the latter are obtained after setting
the initial value of the radiation energy density inside the sphere to
$J=b$ and to $J=b\,R^2/r^2$ outside, where we simply choose $b=1$. The
radial momentum density is instead set to $H_r=0.5\,b\,(R/r)^2$ outside
the sphere and to zero inside. The radiation evolution equation are then
evolved until the system reaches stationarity, which is then compared
with the analytic solutions. From Fig. \ref{fig:rad_sphere} it is evident
that \texttt{FRAC} reproduces the correct result very accurately for the
cases with higher opacity (red and blue). Once again, it is important to
underline that a solution in the case of very high opacity can be
obtained reliably and despite the very sharp change at the surface, only
thanks to the use of the IMEX scheme introduced in
Sec. \ref{sec:implicit} to treat the stiff source terms; also important
are the flux corrections discussed in Sec. \ref{sec:fluxes}, which ensure
the correct fluxes also in the limit of high opacity.

At the same time, Fig. \ref{fig:rad_sphere} also shows that the M1 scheme
fails to accurately reproduce the analytic solution for smaller values of
$\kappa_a$ (green curve). Indeed, while the exterior tail of the energy
density is always computed accurately, this is not the case for the
interior of the sphere for $\kappa_a \lesssim 1$. This error is due to
the closure relation, which gives the correct second moment in the
free-streaming regime and for high optical depths. In the intermediate
regime, however, the analytic closure does not give the correct second
moment (see also Fig. 2 in \citealt{Murchikova2017}). The case of
$\kappa_a=0.125$ falls exactly in this intermediate regime, while the
other two cases (red and blue) do not, which explains the discrepancy in
Fig. \ref{fig:rad_sphere}. Finally, we note that although this test gives
a spherically symmetric result, we still perform the simulations in
3D. As already remarked in \citet{Radice2013}, when using Cartesian
coordinates, the fluxes will propagate across grid cells also in the
angular directions, so that only a 3D simulation is able to reproduce the
correct solution.

\newpage

\subsection{Radiation wave in scattering regime}
\label{sec:scattering}

After having successfully tested free-streaming, absorption and radiation
emission, we next show that also the scattering regime -- the dominating
process inside the dense core of a hypermassive neutron star -- is
reproduced correctly. We recall that scattering is governed by the
coefficient $\kappa_s$, which we here set to a constant value throughout
the domain, while $\kappa_a$ and $\eta$ are set to zero. Scattering
becomes important in the diffusion limit, \ie for very large optical
depths. We here perform the diffusion-wave test from \citet{Pons2000},
which provides an analytic solution of the diffusion equation for
radiation scattering in a homogeneous medium. Starting from an initial
point-like radiation pulse, the solution is given by
\begin{align} 
	J(t,r) &= \left(\frac{\kappa_s}{t}\right)^{d/2}\, \label{eq:gauss1}
	\mathrm{exp}\left(\frac{-3\kappa_s r^2}{4t}\right)\, , \\ \label{eq:gauss2}
	H(t,r) &= \frac{r}{2t} J \,,
\end{align}
with $d$ denoting the number of dimensions (hereafter $d=2$). We have
performed two distinct simulations with $\kappa_s=100$ and
$\kappa_s=10^5$, respectively, on a square grid in Cartesian coordinates
with $x,y\in [-0.5, 0.5]$ and $\Delta x=\Delta y=0.01$. In order to avoid
the divergence at $t=0$ we initialise the simulations according to
Eqs. (\ref{eq:gauss1}) and (\ref{eq:gauss2}) at $t=1$ and $t=200$,
respectively.

Figure \ref{fig:scattering} presents a comparison between the numerical
results (crosses) and the analytic solution (solid lines). For
$\kappa_s=100$ (left panels) we see very good agreement and find the
numerical solution to diffuse only slightly faster. This difference can
be attributed to the additional diffusion intrinsic to our grid-based
code and is reduced with increasing resolution.

The case for $\kappa_s=10^5$, on the other hand, deserves special
attention. The pressure tensor for $\kappa \rightarrow \infty$, in fact,
is given by Eq. (\ref{eq:thick_1}), which is implemented as the limiting
case of the M1 closure. Despite this being the correct pressure in the
diffusion limit, it is known (see, \eg \citealt{Pons2000,
  Oconnor2015}) that the M1 scheme can not correctly reproduce the
diffusion equation in the limit of high optical depths (as is the case
for $\kappa_s=10^5)$. This is most
easily seen from the flux terms in Eqs. (\ref{eq:evolutionrad_1}) and
(\ref{eq:evolutionrad_2}), which have first-order spatial derivatives and
not the second-order derivatives that are expected in a diffusion
equation. It is therefore crucial to correct these fluxes as outlined in
Sec. \ref{sec:fluxes}, where Eq. (\ref{eq:F_asym}) gives the correct flux
in the diffusion limit. In our simulation with $\kappa_s=10^5$, the flux
is dominated [$a=0.01$ in Eq. \eqref{eq:a}] by the correction term in
Eq. (\ref{eq:fluxcorrection}). The difference with the analytic solution
is then a combination of the natural diffusion in a grid-based code and
the flux in Eq. (\ref{eq:evolutionrad_1}), which contributes $\sim 1\%$
to the total flux. In addition to these flux corrections, also the IMEX
scheme is necessary to achieve numerical stability for such a large value
of $\kappa_s$ without having to use a prohibitively small timestep (we
use a Courant-Friedrichs-Lewy (CFL) coefficient of $0.25$ in both
simulations).

As a concluding remark we note that -- except for $\kappa_s\rightarrow
0$, for which Eqs. \eqref{eq:gauss1} and \eqref{eq:gauss2} are no longer
the correct solutions and the problem becomes closer to the one studied
in Sec.  \ref{sec:radiationwave} -- we find similarly good agreement for
all values of $\kappa_s$; once again: this is possible only when using
both the flux corrections and an IMEX scheme.

%
\subsection{Fluid-radiation coupling test}
\label{sec:fluid-coupling}
%
\begin{figure}
\begin{center}
  \includegraphics[width=0.995\columnwidth]{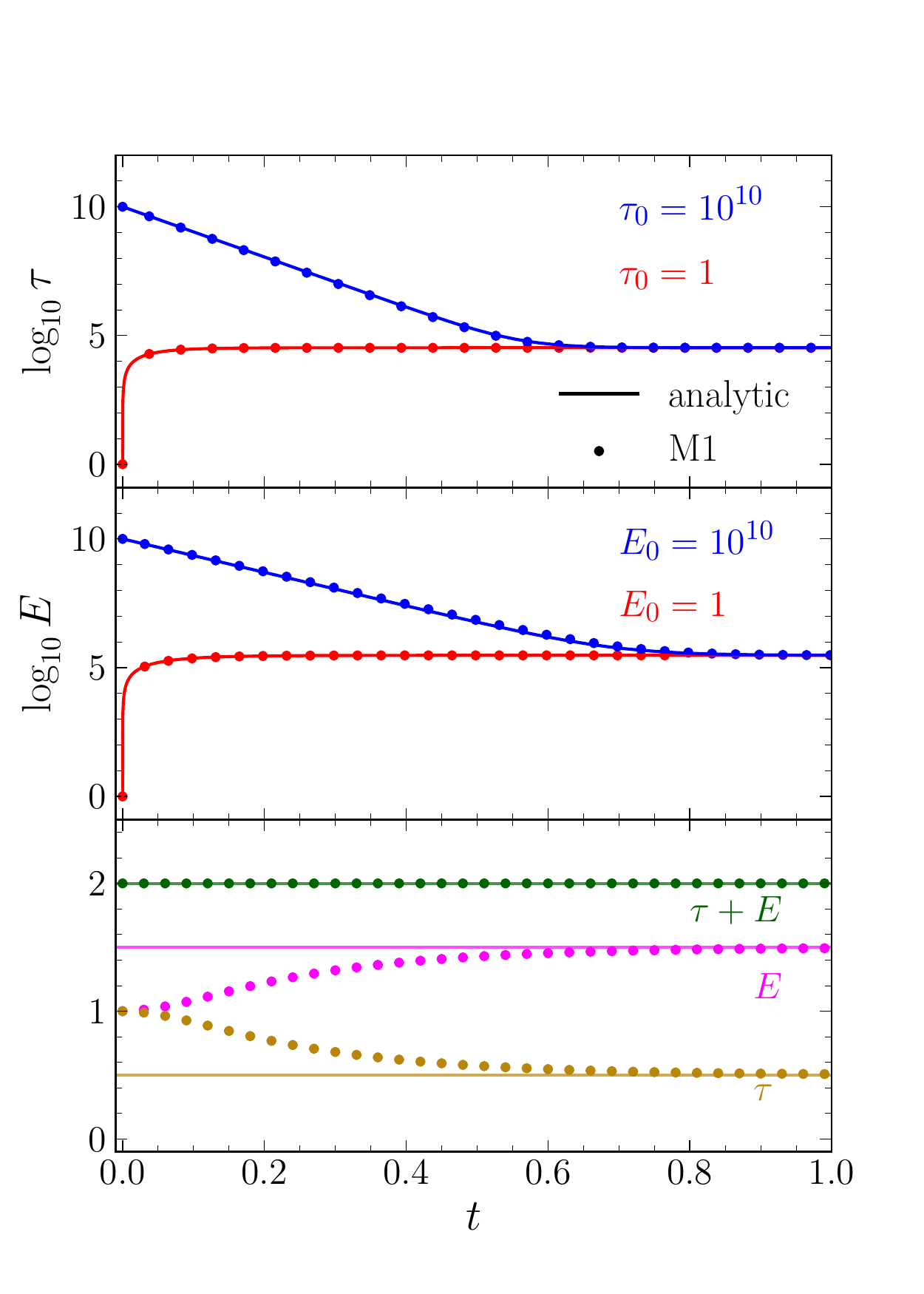}
    \caption{\textit{Top}: Comparison of the analytic and numerical
      solutions for the equilibration process of the rescaled total fluid
      energy density $\tau$ for initial conditions $\tau_0=1$ (red) and
      $\tau_0=10^{10}$ (blue), respectively [\cf Eq.
        (\ref{eq:couplingtest_2})]. \textit{Middle}: Same as above, but
      for the equilibration process of the radiation energy density $E$
      [\cf Eq. (\ref{eq:couplingtest_1})]. \textit{Bottom}: Comparison of
      the analytic and numerical solution of the coupled system of
      Eqs. (\ref{eq:couplingtest_2}) and (\ref{eq:couplingtest_1}). Shown
      with magenta and golden colours are respectively $E$ and $\tau$,
      while green shows the constant total energy, \ie $\tau + E$.}
  \label{fig:fluid_coupling}
\end{center}
\end{figure}

The previous tests only considered the dynamics of the radiation alone,
but not the coupling to an ordinary fluid. A simple test which considers
this coupling is presented in \citet{Turner2001} and
\citet{MelonFuksman2019} and consists of removing the spatial dependence
of the evolution equations
(\ref{eq:evolutionrad_1})--(\ref{eq:evolutionrad_2}) and
(\ref{eq:evolutionfluid_1_rad})--(\ref{eq:evolutionfluid_2_rad}) by
either setting to zero the advection terms or by setting initially
$F_i=0=v_i$, and with $\tau$ and $E$ equal to some spatially homogeneous value. 
In this way, (and neglecting magnetic fields) the evolution equations simplify to
\begin{align}
	\partial_t\, \tau = -G_0 = -\eta + \kappa_a E\,,
	\label{eq:couplingtest_2} \\
	\partial_t\, E = G_0 = \eta - \kappa_a E\,,
        \label{eq:couplingtest_1}
\end{align}
where for the second equalities we have used Eq. (\ref{eq:Smu}) and the
fact that $J=E$ and $H^{\mu}=F^{\mu}$ according to
Eqs. (\ref{eq:fluid2euler_1})--(\ref{eq:fluid2euler_2}) for $v^i=0$. In a
physically realistic setup, the parameters $\eta$ and $\kappa_a$ would be
complex functions of the fluid variables.

For an actual testing of the coupling between the ordinary fluid and the
radiation, we choose a particularly simple (and unphysical) form for
these parameters, namely, $\kappa_a=\rm{const.}$ and $\eta = \kappa_a
\tilde{\eta} \tau$, with $\tilde{\eta}$ set to be a constant. When the
radiation energy density $E$ is held constant over time, the solution of
Eq. (\ref{eq:couplingtest_2}) is then
\begin{equation}
	\tau(t) = \left(\tau_0-\frac{E}{\tilde{\eta}}\right) \exp\left({-\kappa_a
	\tilde{\eta} t}\right) + \frac{E}{\tilde{\eta}}
	\label{eq:coupling_solution_2}\,,
\end{equation}
where $\tau_0:=\tau(t=0)$. Similarly, for a temporally constant
rescaled total fluid energy density $\tau$, the solution of Eq.
(\ref{eq:couplingtest_1}) is
\begin{equation}
  E(t) = \left(E_0-\tilde{\eta}\tau \right) \exp\left({-\kappa_a t}\right) +
  \tilde{\eta} \tau \,,
  \label{eq:coupling_solution_1}
\end{equation}
with $E_0:=E(t=0)$.

The top panel of Fig. \ref{fig:fluid_coupling} compares the numerical
solution of Eq. (\ref{eq:couplingtest_2}), which we obtained after
coupling \texttt{FRAC} with \texttt{BHAC} in two spatial dimensions, with
the corresponding analytic solution \eqref{eq:coupling_solution_2}
relative to two different initial conditions, \ie $\tau_0 = 1$ (red) and
$\tau_0=10^{10}$ (blue), respectively. The radiation energy density is
held constant at $E=10^5$, so that we can test both a fluid- or a
radiation- dominated scenario. Furthermore, we set $\kappa_a=1$ and
$\tilde{\eta}=3$, so that the final equilibrium value (\ie for $G_0 \to
0$) is $\tau_{\rm fin}=\frac{1}{3}\times 10^5$ [\cf
  Eq. (\ref{eq:coupling_solution_2})], which agrees well with the
numerical results. The middle panel in Fig. \ref{fig:fluid_coupling}
compares instead the numerical solution of Eq. (\ref{eq:couplingtest_1})
with the corresponding analytic solution \eqref{eq:coupling_solution_1}
for a constant $\tau=10^5$ and the initial conditions $E_0=1$ (red) and
$E_0=10^{10}$ (blue), respectively. The coefficients $\kappa_a$ and
$\tilde{\eta}$ are set to be the same as before, so that the radiation
energy density should equilibrate to a value of $E_{\rm fin}=3\times
10^5$ [\cf Eq. (\ref{eq:coupling_solution_1})], which is again in good
agreement with the numerical simulations.

In the above tests, energy is constantly injected or removed from the
system via holding $E$ or $\tau$ constant. If both quantities are instead
evolved dynamically, the total energy density of the system, \ie
$\tau+E$, should nevertheless remain constant, independently of the
values chosen for $\kappa_a$ and $\tilde{\eta}$. Considering the case in
which we set $\tau=1=E$ initially, we report with filled circles in the
bottom panel of Fig. \ref{fig:fluid_coupling}, the evolution of $\tau$
(gold), $E$ (magenta) and $\tau+E$ (green). The asymptotic values for the
two energy densities can then be computed from the conditions
$G_0(t\rightarrow\infty)=0$, so that $\eta + \kappa_a E_{\rm fin} = 0$
and $E_{\rm fin} + \tau_{\rm fin} = 2$. Taking $\eta=\kappa_a
\tilde{\eta}\tau_{\rm fin}$, it follows that $E_{\rm
  fin}=2\tilde{\eta}/(1+\tilde{\eta})=3/2$ and $\tau_{\rm
  fin}=2/(1+\tilde{\eta})=1/2$. Clearly, the numerical solution matches
very well the expected equilibrium state.

\section{General-relativistic tests}
\label{sec:grtests}

In what follows, we move away from special relativity, and hence flat
spacetimes, to consider radiation propagation and radiation/fluid
interaction in curved but fixed spacetimes.

\subsection{Curved-beam test}
\label{sec:curvedbeam}

\begin{figure}
\begin{center}
  \includegraphics[width=0.995\columnwidth]{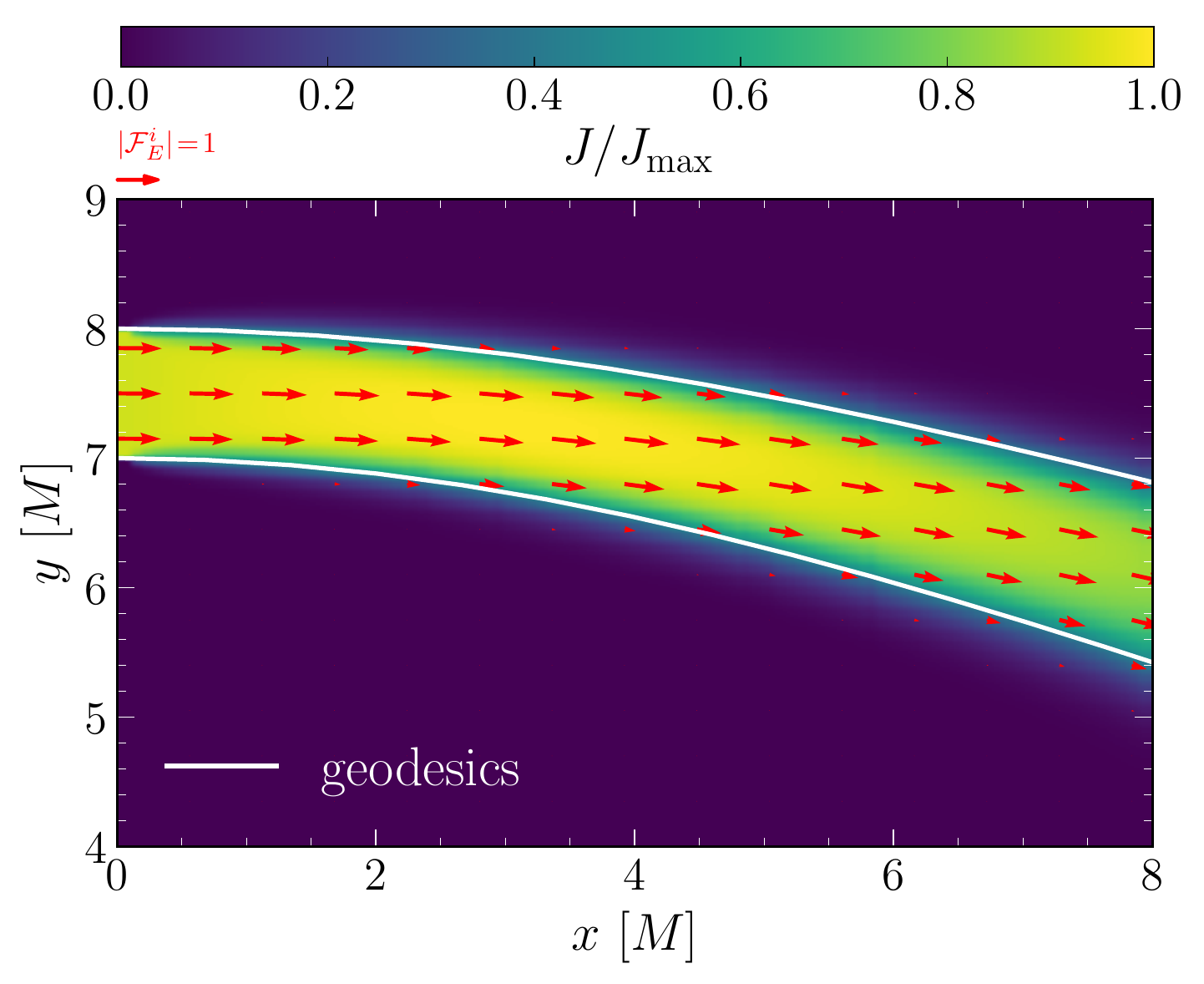}
    \caption{Same as Fig. \ref{fig:beam_straight} but for a radiation
      beam injected in the curved spacetime of a black hole located at
      $(x,y)=0$. The white lines show the corresponding geodesics emitted
      at the edges of the beam.}
  \label{fig:curved_beam_far}
\end{center}
\end{figure}%
\begin{figure}
\begin{center}
  \includegraphics[width=0.995\columnwidth]{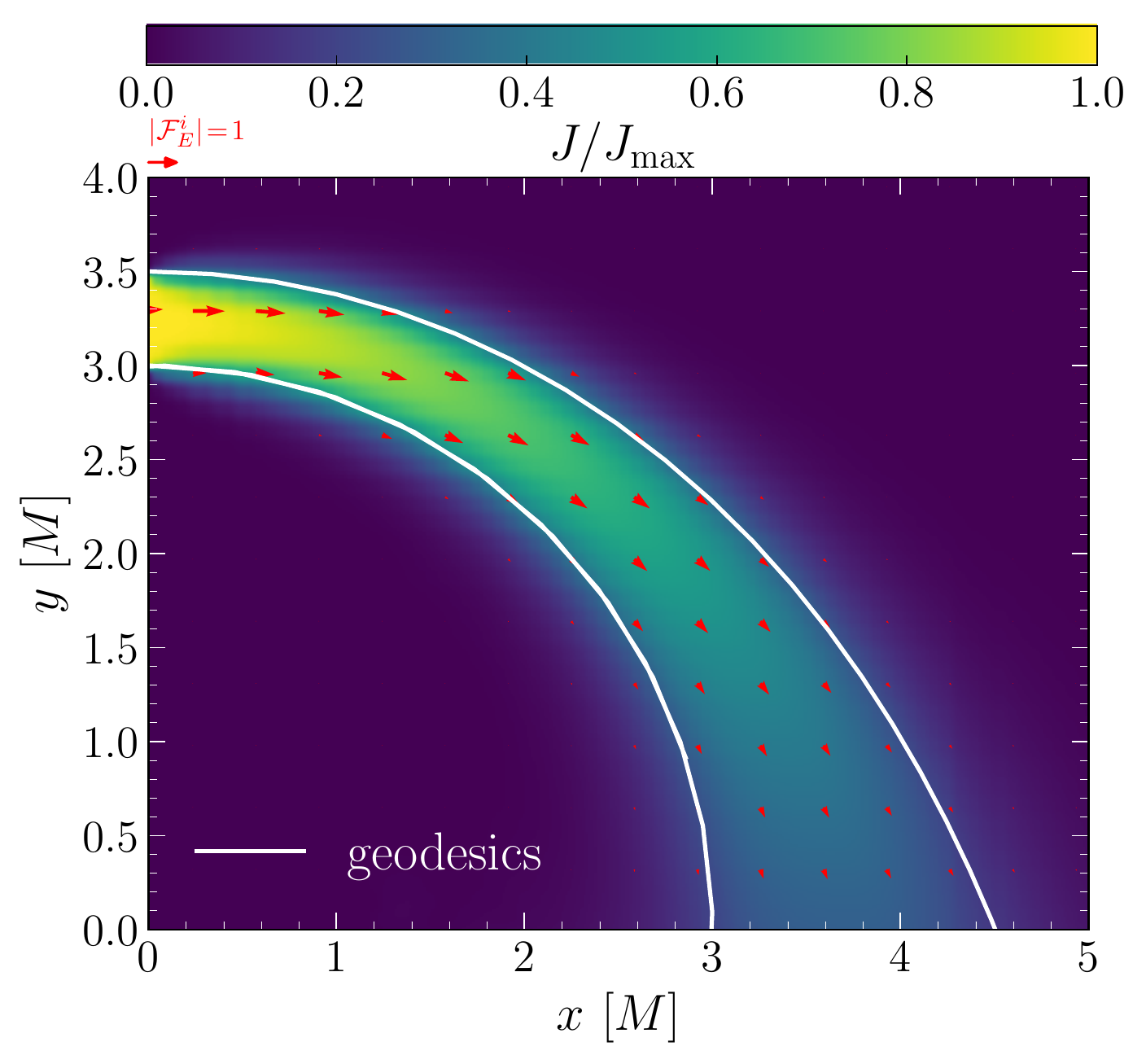}
	\caption{Same as Fig. \ref{fig:curved_beam_far}, but for a beam
          much closer to the black hole and whose lower edge coincides
          with the black hole photon ring.}
  \label{fig:curved_beam_close}
\end{center}
\end{figure}%

As a first test of \texttt{FRAC} in a curved spacetime, we consider the
straight-beam test from Sec.  \ref{sec:beamtest} but within a
Schwarzschild black-hole spacetime of mass $M$, whose metric is expressed
in Cartesian Kerr-Schild coordinates \citep[see,
  \eg][]{Rezzolla_book:2013}. We note that although derivatives of the
metric -- which are needed on the right-hand-sides of
Eqs. \eqref{eq:evolutionrad_1} and \eqref{eq:evolutionrad_2} -- can in
this case be computed analytically, we decide to compute them here
numerically using a fourth-order centred finite-difference scheme.
This choice is only slightly more expensive, but provides us with a much
more general approach and thus with the ability of coupling \texttt{FRAC}
with any GRMHD code in which the spacetime is also dynamical.

As initial data, we set $J=1$ in a region defined by $x<2\Delta x$, where
$\Delta x$ is the grid-spacing in the $x$-direction, and
$y_{\rm{lo}}<y<y_{\rm{hi}}$. Everywhere else we set $J=10^{-15}$. The
momentum density is computed from the condition for the optically thin
limit
\begin{align}
  H^i H_i &= J^2\,, \\
  \mathcal{F}^y_E &= 0 = \mathcal{F}^z_E \,.
\end{align}
This setup ensures that only $\mathcal{F}^x_E$ is nonzero and thus that
the beam is shot into the grid from the left boundary and travels at the
speed of light parallel to the grid's $x$-axis.

Figures \ref{fig:curved_beam_far} and \ref{fig:curved_beam_close} show
the dynamics of the beam, which is obviously no longer straight, as it
is curved by the central black hole located at $(x,y) = 0$. More
specifically, Fig. \ref{fig:curved_beam_far} refers to a beam shot at a
certain distance from the black hole, \ie within a vertical range
$y_{\rm{lo}}=7\,M$, $y_{\rm{hi}}=8\,M$, while the beam in
Fig. \ref{fig:curved_beam_close} is much closer, \ie $y_{\rm{lo}}=3\,M$,
$y_{\rm{hi}}=3.5\,M$, so that the lower edge is actually on the black
hole photon ring.

The trajectory of the beam is compared with the corresponding geodesics
propagating in the same direction and emanating from the vertical edges of
the beam (white solid lines). Clearly, the trajectory of the beams in
both figures is in good agreement with what is expected from the geodesic
motion, but also a certain amount of diffusion is present, as already
encountered in the beam tests in flat spacetime. Note that this
diffusion is more severe for the beam tangent to the photon sphere, since
in this case the beam is highly lensed \citep{EHT_M87_PaperV}. More
importantly, however, the initial and final energies (\ie the energy on
the $y$-axis and that on the $x$-axis in Fig. \ref{fig:curved_beam_close}
differ only by $11.9 \%$.

\subsection{Radiative Michel solution}
\label{sec:bondi}

As a test that involves all terms in the evolution of the
radiation variables and allows for a non-trivial coupling of
\texttt{FRAC}, which the GRMHD code \texttt{BHAC}, we next consider the
problem of spherical accretion onto a nonrotating black hole. In the
absence of radiation and magnetic fields, this ``classic'' problem has
been first analysed by \citet{Bondi52} in Newtonian physics and later in
a general-relativistic context by \citet{Michel72}. Since a realistic
scenario actually involves also radiation and magnetic fields, the
problem of spherical accretion onto a nonrotating black hole has been
explored in many other works, which have either employed simplified
approaches \citep[see, \eg][]{Vitello1978, Begelman1978, Gillman1980} or
fully self-consistent radiative transport using a moment scheme
\citep{Nobili1991,Zampieri1996,Fragile2012,Roedig2012, Sadowski2013,
Fragile2014, McKinney2014}.

Particularly useful among these calculations of the radiative Michel
solution are those of \citet{Nobili1991} and \citet{Zampieri1996}, since
they are the only ones that include the contributions coming from
Comptonization. On the other hand, \citet{Fragile2012} and
\citet{Roedig2012}, have treated the problem assuming the fluid to be
optically thick everywhere, while \citet{Sadowski2013,Fragile2014,McKinney2014}
have made use of the Levermore closure [\cf
  Eq. (\ref{eq:levermore})], which allows to treat both optically thick
and thin regions correctly. In particular, \citet{Sadowski2013,Fragile2014} have
shown that using such a closure, they were able to obtain results far
away from the black hole that were more accurate than those reported by
\citet{Fragile2012}. We here expect a similar accuracy making use of the
Minerbo closure [\cf Eq. (\ref{eq:minerbo})] that is equally effective in
treating the two extreme regimes.

\begin{figure}
\begin{center}
  \includegraphics[width=0.995\columnwidth]{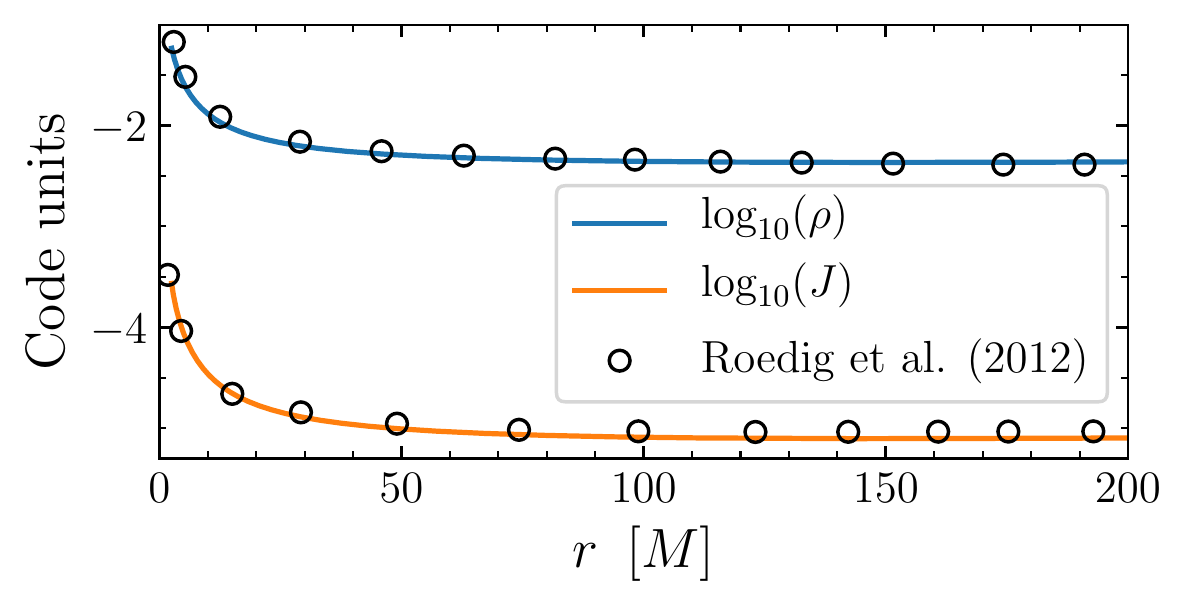}
	\caption{Spherically symmetric accretion onto a Schwarzschild
          black hole with uniform absorption and scattering
          ($\kappa_a=10^{15}$ and $\kappa_s=0$). Shown in blue and orange
          are respectively the fluid rest-mass density and the radiation
          energy density in the fluid frame, while black circles are the
          results taken from Fig. 3 in \citet{Roedig2012}.}
  \label{fig:roedig}
\end{center}
\end{figure}%

\vspace{0.3cm}

\subsubsection{Uniform absorption}

As an initial setup, we adopt the one described by \citet{Roedig2012} and
in which the opacities are assumed to be constant and given by
$\kappa_a=10^{15}$ and $\kappa_s=0$. This scenario is unrealistic as it
lacks a consistent description of the microphysics, but it is useful to
verify that the implementation of all the parts that have been tested
separately in the previous tests, gives the correct results also when the
complete set of equations is employed. Furthermore, what this setup lacks
in terms of physical realism, it makes up for in terms of computational
difficulty. The choice of such a high value for the absorption
coefficient $\kappa_a$, in fact, does represent a severe test of the IMEX
scheme and, as already pointed out by \citet{Roedig2012}, a CFL
coefficient of 0.2 (as used here) would not allow to use $\kappa_a
\gtrsim 1.0$ in a standard explicit scheme.

Following therefore \citet{Roedig2012}, we assume the radiation field to
be that of a black body and set the emissivity accordingly to
\begin{equation}
\label{eq:blackbody}
	\eta=4\pi \, \kappa_a \frac{\sigma_{_{\rm{SB}}}}{c} T^4\,,
\end{equation}
where $\sigma_{_{\rm{SB}}}$ is the Stefan-Boltzmann constant and $T$ the
fluid temperature. For this test we use the same unit system as in
\citet{Zanotti2011}, so that we have a numerical value of
$\sigma_{_{\rm{SB}}}=0.0479$ in our code\footnote{ Note that this unit
  system is different to the one that we use in the following sections
  and which is reported in Appendix \ref{sec:appendixA}, where we
  implement physically realistic microphysics.}.  Although our Minerbo
closure is different from that considered by \citet{Roedig2012}, who use
the Eddington approximation following Eq. (\ref{eq:Lij}), setting
$\kappa_a=10^{15}$ everywhere ensures that only the optically thick limit
is simulated, in which case the two closures are equivalent.

For the same reasons, we consider a black hole with mass $M=2.5\,M_\odot$
and a perfect fluid obeying an ideal-fluid equation of state with
adiabatic index of $\gamma=4/3$.  Furthermore, as in \citet{Roedig2012},
we carry out the evolution of the fluid quantities in one dimension and
using a radial grid in Boyer-Lindquist coordinates ranging from $2.5 <
r/M < 200$, which is covered uniformly with $300$ grid points. The
comparison with the results of \citet{Roedig2012} (empty circles) are
reported in Fig. \ref{fig:roedig} and show a very good agreement both
close to and far away from the black hole; very similar results were
obtained when repeating the calculations in two spatial `dimensions.

\begin{figure*}
\begin{center}
  \includegraphics[width=0.85\textwidth]{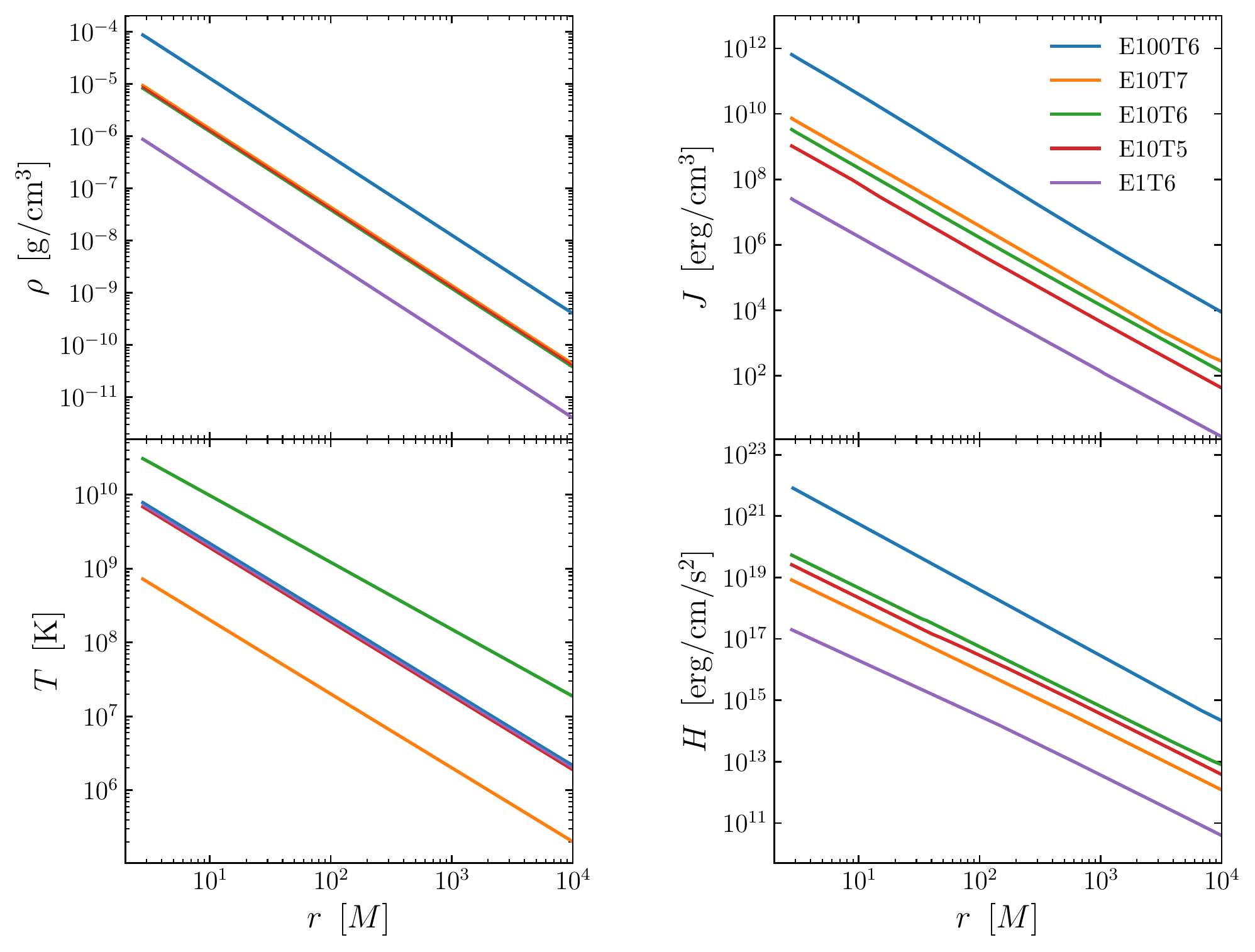}
	\caption{Equilibrium solution for spherically symmetric accretion
          onto a nonrotating black hole. Shown from left to right and top
          to bottom are: the fluid rest-mass density, the radiation
          energy density in the fluid frame, the fluid temperature and
          the radiation energy flux. Different colours correspond to
          different models as described in Table \ref{tab:bondi_runs}.}
  \label{fig:bondi_runs}
\end{center}
\end{figure*}%

\subsubsection{Variable absorption}

We next consider a more realistic setup for the simulation of the
spherical accretion onto a black hole, following the prescription
presented by \citet{Sadowski2013}. In this setup, we take into
account absorption via thermal bremsstrahlung contributing an energy-averaged 
absorption opacity given by \citep{Rybicki_Lightman1986}
\begin{equation}
  \kappa_a = 6.1\times 10^{22}\, T_{\rm{K}}^{-3.5} \rho_{\rm{cgs}}
  \ \rm{cm}^{-1}\,,
\end{equation}
where the temperature is assumed to be in Kelvin and the fluid rest-mass
density in ${\rm g/cm}^3$ (see also the Appendix \ref{sec:appendixA} for
our choice of units). The emission of photons is again treated via
black-body radiation given by Eq. (\ref{eq:blackbody}). We also consider
Thomson scattering, which contributes an energy-averaged scattering opacity
given by
\begin{equation}
	\kappa_s = 0.4\,\rho_{\rm{cgs}} \ \rm{cm}^{-1}\,.
\end{equation}
The simulation is initialised by setting the fluid density as the
free-fall density given by
\begin{equation}
	\rho = \frac{\dot{M}}{{4\pi r^2 v}}\,,
\end{equation}
where $\dot{M}$ is the accretion rate and $v= \sqrt{2M/r}$ is the modulus
of the fluid three velocity, \ie $v^2=v^i v_i$.
The components of the three-velocity are instead given by 
\begin{align}
	v^r &= -\sqrt{v^2/g_{rr}}\,, \qquad v^\phi = 0 = v^\theta \,,
\end{align}
where $g_{rr}$ is the radial component of the four-metric. Furthermore,
we specify the temperature $T_0$ at some fiducial radius
$r_0$ as a free
parameter. We assume a perfect fluid and a polytropic equation of state
of the form $P\propto \rho^\gamma$ with the adiabatic index $\gamma$. The
fluid initial pressure can then be computed as
\begin{equation}
  P= \frac{k_{_{\rm B}}\, T_0}{\mu m_p} \frac{\rho^\gamma}{\rho_0^{1-\gamma}} \,,
\end{equation}
where $k_{_{\rm B}}$ is the Boltzmann constant, $m_p$ the proton mass,
$\rho_0 := \rho(r=r_0)$, and $\mu$ the mean molecular weight, which is
given by $\mu=0.5$ for fully ionised hydrogen (\citet{Fragile2012} and
\citet{McKinney2014} have a similar setup, but initialise the temperature
rather than the pressure). The numerical values of the variables depends
on the choice of units and is reported in Appendix \ref{sec:appendixA}
for completeness. Finally, we choose the adiabatic index as
\begin{equation}
	\gamma = 1 + \frac{2}{3}\left(\frac{\beta_{\rm rad}+1}{\beta_{\rm rad}+2}\right)\,,
\end{equation}
where $\beta_{\rm rad}$ is the ratio of the fluid-to-radiation pressure
at the initial time and the adiabatic index is constrained to be $4/3 <
\gamma < 5/3$. The radiation energy density is initialised as
\begin{equation}
E = 3 P/\beta_{\rm rad}\,,
\end{equation}
and the radiation fluxes are set to $F_i = 0$. We have verified that
using somewhat different initial conditions still leads to the same
equilibrium state.

In summary, the setup presented here to simulate spherical accretion onto
a black hole has five free parameters: the
black hole mass $M$, the accretion rate $\dot{M}$, the temperature $T_0$,
the ratio of fluid-to-radiation pressure $\beta_{\rm rad}$, and the
matching radius $r_0$. Hereafter we will hold fixed: $r_0=2\times
10^4\,M$ and $M=3M_\odot$, while $T_0$, $\dot{M}$ and $\beta_{\rm
  rad}$ are varied as described in Tab. \ref{tab:bondi_runs} (\cf Tab. 5
in \citealt{Sadowski2013}).
\begin{table}
  \begin{center}
  \begin{tabular}{ l c c c c }
	Model & $\dot{M}/\dot{M}_{_{\rm{Edd}}}$ & $T_0$ & $1/\beta_{\rm rad}$ & $L/L_{_{\rm{Edd}}}$ \\
  \hline
	E1T6 & 1.0 & $10^6$ & $1.2\times 10^{-4}$ & $2.33\times 10^{-8}$ \\
	E10T5 & 10.0 & $10^5$ & $1.2\times 10^{-7}$ & $6.62\times 10^{-7}$ \\
	E10T6 & 10.0 & $10^6$ & $1.2\times 10^{-4}$ & $2.65\times 10^{-6}$ \\
	E10T7 & 10.0 & $10^7$ & $1.2\times 10^{-1}$ & $6.41\times 10^{-6}$ \\
	E100T6 & 100.0 & $10^6$ & $1.2\times 10^{-4}$ & $2.01\times 10^{-4}$ \\
  \hline
\end{tabular}
\caption{Parameters for the different runs following \citet{Fragile2012,
    Sadowski2013} and the final luminosities extracted at $r=1000M$. The
  accretion rates/luminosities are reported as multiples of the Eddington
  accretion rate/luminosity.}
\label{tab:bondi_runs}
  \end{center}
\end{table}

Although the spherical symmetry would allow for one-dimensional
simulations, we still use two dimensions in order to test as many terms
in our code as possible; however, we have verified that the final results
are independent of the dimensionality chosen for the simulations. We
employ \texttt{BHAC} on a two-dimensional grid covered with modified
Kerr-Schild spherical polar coordinates as described in
\citet{Porth2017}. The radial grid ranges from $1.1\,r_S$ to $10^4\,r_S$,
where $r_S=2M$ is the Schwarzschild radius, and employs $600$ grid-points
that are equally spaced in the underlying coordinate system, which itself
uses a logarithmic radial coordinate. The angular grid, instead, ranges
from $0$ to $\pi/2$ and uses $40$ grid-points. Outflow boundary
conditions are used at the outer edge of the computational domain.

The luminosity is computed as $L=4\pi r^2 F$, where $F:=\sqrt{F_{\mu}
  F^{\mu}}$, and is extracted at $r=10^3\,M$; our results change only
marginally when extracting the luminosity at somewhat larger or smaller
radii. Also, while the bolometric luminosity should be computed from the
radiation flux in the Eulerian frame, we find the same results when
computing the flux in the comoving frame instead. This is because far
away from the black hole, \ie at $r=10^3\,M$, the fluid is almost static,
so that $F\sim H$.

\begin{figure}
\begin{center}
  \includegraphics[width=0.995\columnwidth]{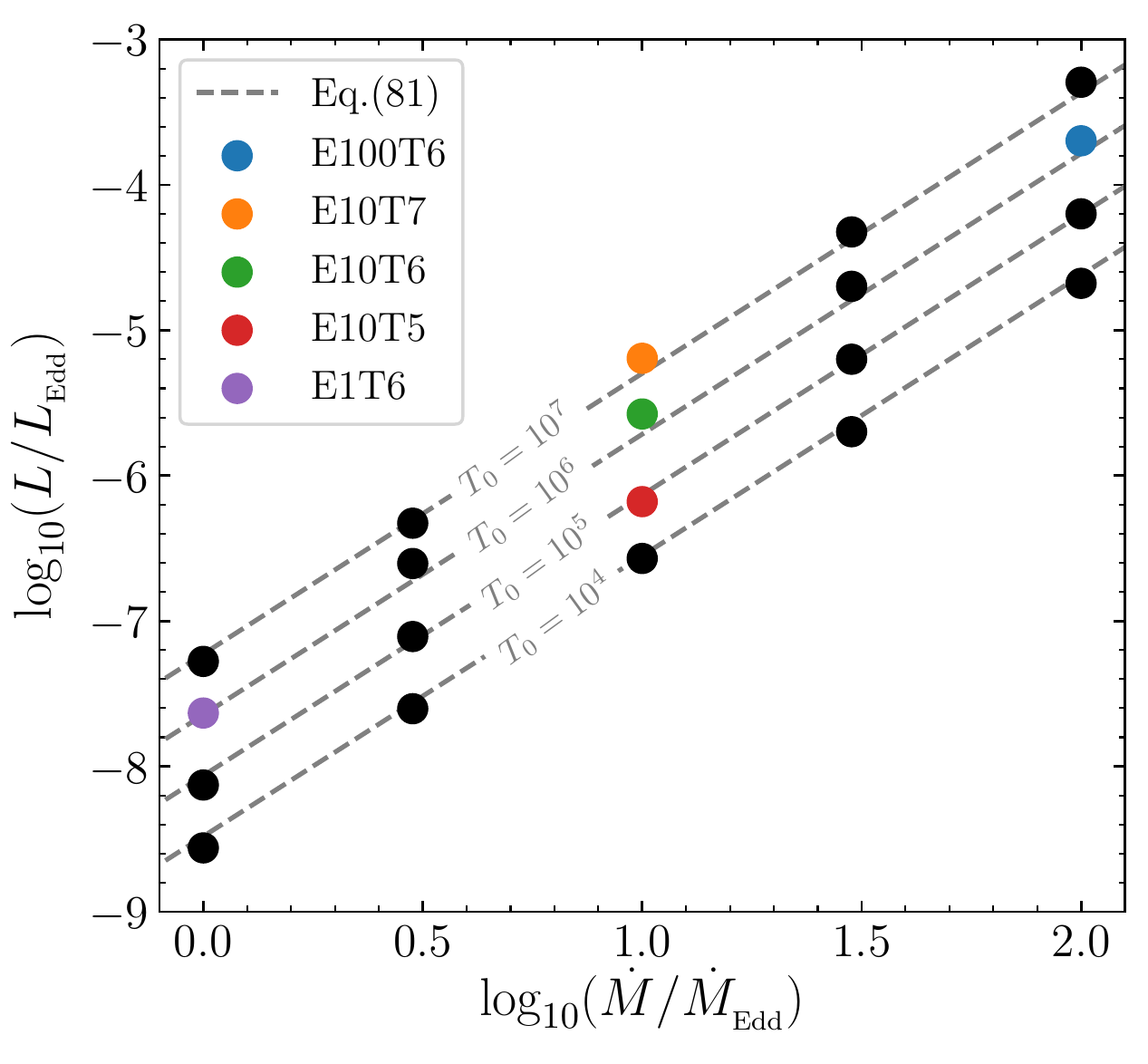}
	\caption{Rescaled luminosity $\mathcal{L}:={L}/{L}_{_{\rm{Edd}}}$
          as a function of the normalised mass-accretion rate for
          simulations with
          $\dot{\mathcal{M}}:=\dot{M}/\dot{M}_{_{\rm{Edd}}}=1, 3, 10, 30,
          100$ and $T_0=10^4,10^5,10^6,10^7$. Coloured circles show the
          same runs as in Fig. \ref{fig:bondi_runs} and are listed in
          Tab. \ref{tab:bondi_runs}. The dashed lines show the fit using
          Eq. (\ref{eq:fit}).}
  \label{fig:le}
\end{center}
\end{figure}

The results of our simulations are reported in Fig. \ref{fig:bondi_runs},
where the fluid rest-mass density and temperature, together with the
radiation energy density and energy flux, are shown for the final
equilibrium state. Note that all quantities show smooth radial profiles,
in contrast to what was found by \citet{Fragile2012}, where the solution is
smooth only close to the black hole, \ie in the optically thick regime.
This difference was observed already by \citet{Sadowski2013,Fragile2014,
McKinney2014} and, as already mentioned, it is due to the
choice of a better closure relation.

Interestingly, in all cases the radiation energy density (momentum
density) follows a simple power-law in radius of the form $\propto
r^{-a}$ ($\propto r^{-b}$), where $a\sim1.91$ ($b\sim1.78$). We note that
the values reported in Fig. \ref{fig:bondi_runs} are similar but also
systematically smaller by a factor $\sim 4$ than those reported by
\citet{Sadowski2013} and \citet{Fragile2014}. This is due to the fact
that the latter are reported in the so-called ``radiation rest frame'',
that is, the frame in which the radiation fluxes vanish.  We do not make
use of this frame as we have a single frame -- the fluid frame -- and
report all quantities in this frame. However, it is possible to transform
from one frame to the other [see Eqs. (3) and (4) in \citet{Fragile2014}]
and thus compare more closely the two sets of results. In this way, we
find that the differences are much smaller and within the expected
variance among the various codes. In particular, we find good agreement
with \citet{Fragile2014} (within a factor $\lesssim 2$) and a slightly
worse agreement with \citet{Sadowski2013}. However, similar differences
exist even between the results of \citet{Fragile2014} and
\citet{Sadowski2013}, who follow the same implementation and closure
scheme.

At the same time, we do not measure any systematic offset when comparing
our results with those of \citet{Roedig2012} (\cf Fig.
\ref{fig:roedig}), who implement the two-moment scheme following the
exact same approach (despite their treatment of the closure) as we
do. Notwithstanding these small descrepancies, all simulations show the
same overall qualitative behavior and yield quantitative values of the
same order of magnitude as those presented so far in the literature
\citep{Fragile2012,Sadowski2013,Fragile2014,McKinney2014}.

As a final but important side-product of this systematic exploration of
the space of parameters, Fig. \ref{fig:le} reports in the
$({\mathcal{L}}, \dot{\mathcal{M}})$ plane, where
${\mathcal{L}}:=L/L_{_{\rm{Edd}}}$ and $\dot{\mathcal{M}} :=
\dot{M}/\dot{M}_{_{\rm{Edd}}}$ ($L_{_{\rm{Edd}}}$ and $\dot{M}_{_{\rm{Edd}}}$
are the Eddington
luminosity and mass-accretion rate, respectively\footnote{We recall that,
  when writing explicitly all the constants, the Eddington luminosity is
  defined as $L_{\rm Edd} := 4\pi G c M {m_p}/{\sigma_{_{\rm {T},e}}}
  \simeq 1.26\times 10^{38}\, ({M}/{M_\odot})$, while the Eddington
  mass-accretion rate is instead $\dot M_{\rm Edd} := {L_{\rm Edd}}/{c^2}
  \simeq 1.39\times 10^{17}\,({M}/{M_\odot})\, {\rm g \ s^{-1}}$, where
  $\sigma_{_{\rm {T},e}}\simeq 6.65\times 10^{-25}\, {\rm cm}^2$ is the
  Thomson cross-section of electrons.}), the results of 20 different simulations
with mass-accretion rates and temperatures given by $\dot{\mathcal{M}} =
1,3,10,30,100$ and $T_0=10^4, 10^5, 10^6, 10^7\,{\rm K}$. Figure \ref{fig:le} clearly
indicates a linear dependence between $\mathrm{log}_{10}({\mathcal{L}})$ and
$\mathrm{log}_{10}(\dot{\mathcal{M}})$, that we express as (dashed lines)
\begin{equation}
\label{eq:fit}
	\log_{10}\left[\, {\mathcal{L}} (\dot{\mathcal{M}},T_0)\,\right] = a\,
	\log_{10}\left(\dot{\mathcal{M}}\right) + b\,,
\end{equation}
where $a$ and $b$ are two coefficients that are in principle functions of
the temperature, \ie $a=a(T_0)$ and $b=b(T_0)$. In practice, we find $a$ to be
roughly independent of the temperature, \ie $a = (1.930 \pm 0.067)$,
while $b = (0.418 \pm 0.026)\log_{10}(T_0) - (10.154 \pm 0.144)$.
The relation \eqref{eq:fit} can be inverted to find the accretion
efficiency $\varepsilon$
\begin{align}
	\label{eq:efficiency}
	\varepsilon &:= \frac{\mathcal{L}}{\dot{\mathcal{M}}} = 10^{b/a} \, {\mathcal{L}}^{(a-1)/a} \\
	&\approx 5.481\times
	10^{-6}\ T_0^{0.217}\,\left(\frac{L}{L_{_{\rm{Edd}}}}\right)^{0.482}
	\nonumber \\
	&=7.41\times
	10^{-7}\ \left(\frac{T_0}{10^6\,\rm{K}}\right)^{0.22}\left(\frac{L}{L_\odot}\right)^{0.48}\,\left(\frac{M}{M_\odot}\right)^{0.48}
	\nonumber
	\,. 
\end{align}
Expressions \eqref{eq:fit} and \eqref{eq:efficiency} are particularly
useful as they allow to relate simply the observed luminosity with either
the mass of the black hole or the physical properties of the plasma.

\subsection{Perturbed radiative Michel solution}
\label{sec:perturbation}
\begin{figure*}
\begin{center}
  \includegraphics[width=0.98\textwidth]{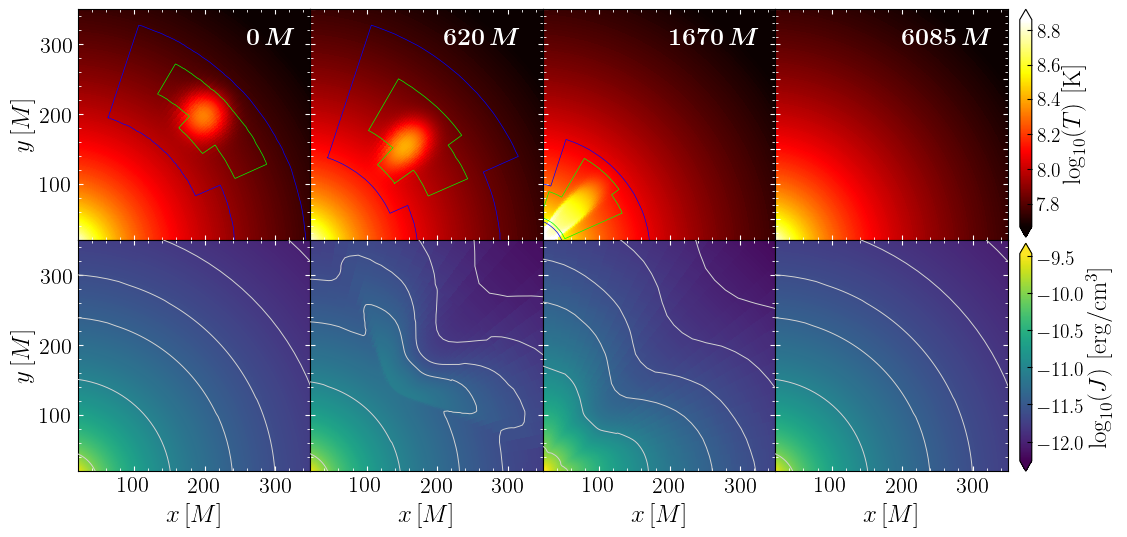}
	\caption{Snapshots of a spherically symmetric accretion onto a
          nonrotating black hole after a Gaussian perturbation in the
          temperature distribution. Shown at four representative times
          are the fluid temperature (top) and the radiation energy
          density in the fluid frame (bottom). The coloured contours in
          the top panels indicate the AMR structure (the asymmetry of
          this structure is due to the misalignment of \texttt{BHAC}'s
          block-based grid with the perturbation). The grey contours in
          the bottom panel show isosurfaces of the radiation energy
          density at
	  $\mathrm{log}_{10}(J)=[-12,-11.8,-11.6,-11.4,-11,-10]$.}
  \label{fig:perturbed1}
\end{center}
\end{figure*}

As a final test and a way to explore the stability properties of the
Michel solution in the presence of a radiation field (see also
\citet{Tejeda2020,Waters2020} for a related exploration in pure
hydrodynamics), we next deviate from spherical symmetry via introducing a
perturbation to the equilibrium solutions of Sec. \ref{sec:bondi}. As a
representative initial background configuration we consider model
$\rm{E10T6}$ and introduce a perturbation in the initial temperature
distribution of the form
\begin{equation}
	\Delta T = A\, \textrm{exp}\left( \frac{(x-x_0)^2}{\sigma^2} +
	\frac{(y-y_0)^2}{\sigma^2}\right)\, ,
\end{equation}
where we set $\sigma=800\,M$, $A=2.5$ and $x_0=y_0=200\,M$ in order to
fix size, amplitude and position of the perturbation, respectively. This
perturbation immediately changes the density and pressure of the
configuration through the equation of state, but does not affect the
initial data of the radiation. The left-most panel in
Fig. \ref{fig:perturbed1} shows the temperature (top) and the unaffected
radiation-energy density (bottom) of this perturbed initial
configuration. The grid extent is the same as before, with the exception
that three levels of AMR are now employed. This is not just useful to
resolve the temperature hot-spot introduced with the perturbation, but
also to test the coupling between \texttt{BHAC}'s AMR routines and
\texttt{FRAC} (see also Sec. \ref{sec:coupling}).
\begin{figure*}
\begin{center}
  \includegraphics[width=0.7\textwidth]{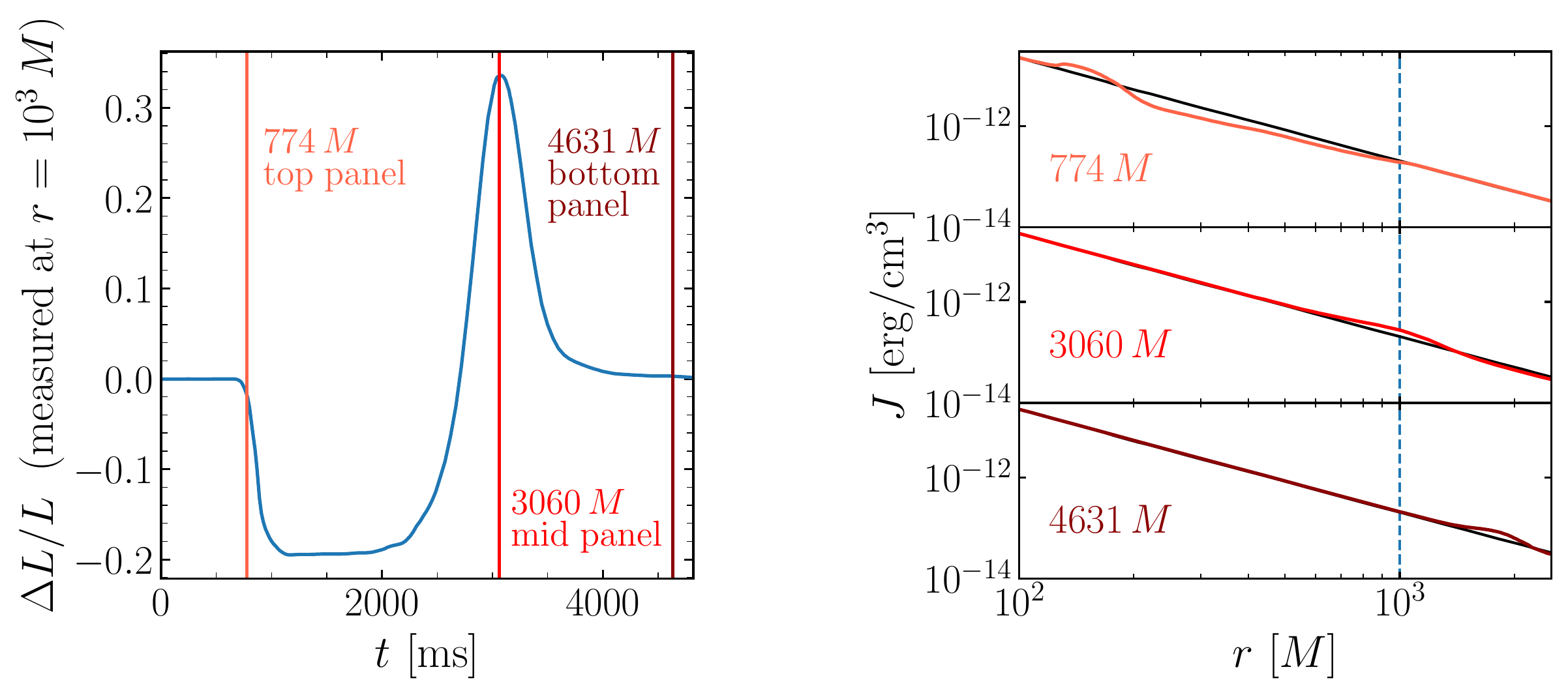}
	\caption{\textit{Left:} Relative difference in the bolometric
          luminosity extracted at $r=10^3\,M$ produced by the temperature
          perturbation. \textit{Right:} Radial cuts of the radiation
          energy density at the three times indicated with vertical
          coloured lines in the left panel. Shown with black lines is the
          background equilibrium solution, while the blue-dashed lines
          mark the distance $r=10^3\,M$ where the luminosity is
          measured. Note that the excess in luminosity at $t=3060\,M$
          results from the enhancement of the perturbation as it
          approaches the black hole (\cf Fig. \ref{fig:perturbed1}).}
  \label{fig:perturbed2}
\end{center}
\end{figure*}%

The overall dynamics of the perturbed accretion problem is reported in
Fig.  \ref{fig:perturbed1}. When starting the simulation, the radiation
field immediately deviates from spherical symmetry (see second
bottom-panel from the left) due to the increased emissivity and opacities
that arise from the increased temperature around the position of the
perturbation. At the same time, the underlying accretion drags the
hot-spot towards the black hole (second top-panel from the left), which
happens independent of the coupling to the radiation. After the hot-spot
plunges into the black hole (third top-panel from left) the fluid returns
to its previous equilibrium (right-most top-panel) unaffected by the
radiation, whose energy is again several orders of magnitudes smaller
than that of the fluid.

The radiation field (bottom panels in Fig. \ref{fig:perturbed1}), on the
other hand, shows a region of increased energy density that falls towards
the black hole and at the same time a region of decreased energy density
that develops behind the hot-spot. The latter region moves radially
outward leading to a decreasing luminosity. The decrease in luminosity
can be seen in the left panel of Fig. \ref{fig:perturbed2}, which shows
the relative difference in the bolometric luminosity with respect to the
steady-state solution as a function of time. The minimum in luminosity
(see $t=1000-2000\,M$) is followed by a sharp increase with the peak
luminosity at $\sim 3060\,M$, when extracting the luminosity at
$r=10^3\,M$. The outward propagation of the perturbation can be tracked
in the right panels in Fig. \ref{fig:perturbed2}, which show radial cuts
of $J$ at an angle of $\theta=\pi/4$ and at three different times. At the
beginning (top panel) and at the position of the perturbation (\ie
$r=200\,M$), the local maximum and minimum in $J$ can be seen
forming. The decrease in $J$ behind the perturbation can be interpreted
as a ``shadowing effect'' introduced by the perturbation and hence is
rather narrow in the angular direction (\cf Fig. \ref{fig:shadow}). Also,
while the deficit in $J$ propagates outwards at the speed of light, the
corresponding increase in $J$ falls towards the black hole. When such
excess in the radiation energy approaches the black hole, a ring of
increased radiation energy density forms around the black hole and a part
of it propagates outwards to infinity (see mid and bottom panels in
Fig. \ref{fig:perturbed2}), while another part is clearly captured by the
black hole. Eventually, the radiation field returns to its initial
equilibrium state. To the best of our knowledge, this is the first
evidence that the radiative Michel solution is nonlinearly stable under
perturbations in the radiation field.

\section{Conclusions}
\label{sec:conclusions}

We have implemented the moment scheme developed by \citet{Thorne1981}
\citep[see also][for numerical
  implementations]{Rezzolla1994,Shibata2011,Cardall2013b}, truncating the
moment expansion at the first two moments, \ie in what is known as the M1
scheme. The closure to the moment expansion is obtained via the Minerbo
closure that -- within the ``grey'' approximation in which the frequency
dependence is integrated away -- provides an accurate description of
radiative transport in the optically thick and thin limits and
a reasonable approximation for the intermediate regime.

Our new radiation code \texttt{FRAC} is logically similar to the one
presented by \citet{Foucart2015a}, but has the important advantage of
making use of an Implicit-Explicit (IMEX) IMEX scheme in order to tackle
the stiffening of radiative-transfer equations in the regimes of very
high opacity. Indeed, adopting this technique is essential to obtain
accurate solutions at acceptable computational costs in those regimes
where the absorption or scattering opacities are very large. To this
scope, we have provided a systematic description of the steps necessary
-- and of the potential pitfalls to be encountered -- when implementing a
two-moment scheme within an IMEX scheme to include radiative-transfer
contributions in numerical simulations of general-relativistic plasmas.

\texttt{FRAC} has been developed as a stand-alone code and can therefore
be coupled to any other code solving the equations of GRMHD, either as on
fixed or on dynamical spacetimes. This feature has allowed us to couple
\texttt{FRAC} to \texttt{BHAC}, a GRMHD code recently developed to
explore accretion processes onto black holes \citep{Porth2017}, and work
is in progress to obtain a similar coupling with \texttt{FIL}
\citep{Most2018b,Most2019b}. Hence, we expect that similar couplings will
be possible with other codes, \eg to those publicly available within the
Einstein toolkit \citep{loeffler_2011_et}, when a public version of
\texttt{FRAC} will be released.

We have shown in a number of tests, in special and general relativity,
that \texttt{FRAC} performs well for all scenarios encountered within the
simulation of accretion problems onto compact objects or the merger of
binary systems of compact objects. The only exception to this successful
suite of tests is represented by the crossing-beam problem, whose
accurate treatment requires a different method than the one employed in
this work.

As a first physically relevant application of the new code, we have
simulated the problem of spherically symmetric accretion onto a
nonrotating black hole, \ie the radiative Michel solution, thus coupling
\texttt{FRAC} with \texttt{BHAC} within an AMR approach. Investigating a
large parameter space, we derived a simple expression [\cf
  Eq. (\ref{eq:efficiency})] that links the black-hole accretion
efficiency to the three properties of the system, namely, the
temperature, the bolometric luminosity and black-hole mass.

We further evaluated this accretion problem away from spherical symmetry
by introducing a Gaussian perturbation in the initial temperature
distribution. We found the system to return to its spherically symmetric
equilibrium, which is achieved by radiating the excess energy to
infinity. This process is captured faithfully in the lightcurve, which
shows first a minimum produced by the shadowing effect introduced by the
perturbation, followed then by a maximum resulting from the accretion of
the perturbation onto the black hole. Because the system eventually
recovers the stationary solution in the absence of a perturbation, this
is, to the best of our knowledge, the first evidence that the radiative
Michel solution is nonlinearly stable under perturbations in the
radiation field.

\section*{Acknowledgements}
It is a pleasure to thank Elias Most, Fabio Bacchini and Bart Ripperda
for useful discussions. LRW acknowledges support from HGS-HIRe. Support
also comes in part from ``PHAROS'', COST Action CA16214; LOEWE-Program in
HIC for FAIR; the ERC Synergy Grant ``BlackHoleCam: Imaging the Event
Horizon of Black Holes'' (Grant No. 610058). The simulations were
performed on the SuperMUC and SuperMUC-NG clusters at the LRZ in
Garching, on the LOEWE cluster in CSC in Frankfurt, and on the HazelHen
cluster at the HLRS in Stuttgart.


\bibliographystyle{mnras}
\input{paper.bbl}


\appendix
\section{Units and units conversions}
\label{sec:appendixA}

Special attention has to be paid to the system of units used when
coupling radiative-transfer and GRMHD codes (see also Appendix A of
\citealt{Rezzolla_book:2013}). Indeed, it is not uncommon to encounter
tedious problems when converting quantities from the units that are
routinely used in the GRMHD codes (normally employing geometrised units)
and the units in which physical quantities -- such as the opacities and
emissivities -- are routinely expressed (normally employing CGS
units). \texttt{BHAC}, for instance, makes use of geometrised units with
$c=1=G$ and it assumes that there is a stationary background metric of
mass $M$, so that all lengthscales can be scaled in terms of such a mass
as $[L] = G\, M / c^2$, times as $[T] = [L] / c$, and velocities as
$[V]=c$. Due to this scale invariance one can typically choose $M=1$ for
convenience.

In the presence of radiation, however, this scale invariance is broken,
because new scales are introduced by microphysical quantities, \eg the
proton mass. While $M=1$ is still a reasonable choice when considering
matter at high densities (such as in simulations of binary neutron
stars), it may lead to rest-mass densities $\mathcal{O}(10^{-22})$ in
typical simulations of accretion problems onto supermassive black holes,
thus exposing the numerical calculations to floating-point errors. To
avoid this problem, we exploit the fact that the accreted mass is much
smaller than that of the central black hole and thus can be neglected as
a contribution for the spacetime curvature, allowing us to define an
independent mass-scale for the fluid. We choose this scale via the
Eddington mass-accretion rate, so that $[M] = \dot{M}_{_{\rm{Edd}}}
\times [T]$ and all related quantities follow from this scaling, \eg the
rest-mass density will have dimensions $[\rho] = [M]^{-2}$. An additional
advantage of this specific system of units is that the accretion rate is
naturally expressed in terms of $\dot{M}_{_{\rm{Edd}}}$ and the computed
luminosity will be already rescaled in terms of the Eddington luminosity.

For easy reference, we report below some useful conversion relations for
the natural constants:
\begin{align}
	k_{_{\rm B}}^{\texttt{code}} &= \frac{k_{_{\rm
	B}}^{\texttt{CGS}}}{[M][V]^2\, \rm{K}}
	  = 2.22992 \times 10^{-49}
	  \left(\frac{M}{M_\odot}\right)^{-2} \rm{K}^{-1}\, , \\
	m_p^{\texttt{code}} &= \frac{m_p^{\texttt{CGS}}}{[M]} = 2.42798\times
	10^{-36} \left(\frac{M}{M_\odot}\right)^{-2}\, , 
\end{align}
where the temperature unit Kelvin, \ie $\rm{K}$, remains unchanged. The
radiation constant $a_{_{\rm R}}\coloneqq 4\sigma_{_{\rm
    SB}}/c=7.5657\times 10^{-15} \rm erg\, cm^{-3}\, K^{-4}$ is
implemented in CGS units and enters in the computation of the
emissivity. For the latter, we first express quantities in CGS units and
then convert to code units via
\begin{equation}
	\eta^\texttt{code} = \eta^\texttt{CGS} \times \left(\frac{[L]^2\,
	[T]^2}{[M]}\right) = 7.67822\times 10^{-13}
	\left(\frac{M}{M_\odot}\right)^2 \eta^\texttt{CGS}
\, .
\end{equation}
Finally, the opacities are converted straightforwardly via
\begin{equation}
	\kappa^\texttt{code} = \kappa^\texttt{CGS} \times [L] = 1.47760 \times
	10^5 \left(\frac{M}{M_\odot}\right) \kappa^\texttt{CGS}
        \,.
\end{equation}

\end{document}